\documentclass[letterpaper,twocolumn,10pt]{article}
\usepackage{usenix-2020-09}

\usepackage{amsmath} 
\usepackage{enumitem}

\usepackage{balance}
\usepackage{tikz}
\usetikzlibrary{arrows}
\usepackage{pgfplots}
\pgfplotsset{compat=1.18}
\definecolor{mygreen}{HTML}{009E73}
\definecolor{myorange}{HTML}{E69F00}
\definecolor{myblue}{HTML}{0072B2}
\definecolor{myred}{HTML}{D55E00}
\usepackage{amsmath}
\usepackage{subcaption}
\usepackage{ifthen}
\usepackage{enumitem}
\usepackage{xspace}
\usepackage{multirow}
\usepackage{tabularx}
\usepackage{booktabs}
\usepackage{hyperref}
\usepackage[capitalise,nameinlink]{cleveref}
\usepackage[bottom]{footmisc}
\usepackage{graphicx}
\usepackage{comment}
\usepackage[utf8]{inputenc}
\usepackage{todonotes}
\usepackage{xurl}
\crefname{section}{\S}{\S{}}
\crefname{figure}{Fig.}{Fig.}
\crefname{table}{Tab.}{Tab.}
\crefname{definition}{Def.}{Def.}
\crefname{proposition}{Prop.}{Prop.}
\crefname{appendix}{Appx.}{Appx.}
\crefname{theorem}{Theo.}{Theo.}
\crefformat{section}{\S#2#1#3}

\newcommand{\usec}{$\mu$s\xspace}
\newcommand{\projname}{SPX\xspace}
\newcommand{\AR}{AllReduce\xspace}
\newcommand{\ATOA}{All2All\xspace}
\newcommand{\AG}{AllGather\xspace}
\newcommand{\SR}{SendReceive\xspace}

\renewcommand{\paragraph}[1]{\noindent\textbf{#1.}}

\newbool{public_version}
\setbool{public_version}{false}

\ifbool{public_version}{
    \newcommand{\ms}[1]{}
    \newcommand{\ag}[1]{}
    \newcommand{\sajy}[1]{}
    \newcommand{\albert}[1]{}
    \newcommand{\rani}[1]{}
}{
    \newcommand{\ms}[1]{{\color{red}\big[MS: #1]}}
    \newcommand{\os}[1]{{\color{blue}\big[OS: #1]}}
    \newcommand{\ag}[1]{{\color{blue}\big[AG: #1]}}
    \newcommand{\sajy}[1]{{\color{green!50!black}\big[Sajy: #1]}}
    \newcommand{\albert}[1]{{\color{purple}\big[Albert: #1]}}
    \newcommand{\rani}[1]{{\color{purple}\big[Rani: #1]}}
}

\usepackage{pgfplots}
\usepgfplotslibrary{statistics}
\pgfplotsset{
  nccl axis/.style={
    width=0.4\textwidth,      
    height=0.28\textwidth,    
    xmin=90000,
    xmax=2e10, 
    ymin=-7.897,
    ymax=191.577,
    axis x line*=bottom,
    axis y line*=left,
    xtick={
      16384, 262144, 4194304, 67108864, 1073741824, 1.7179869184e10
    },
    xticklabels={
      16 KB, 256 KB, 4 MB, 64 MB, 1 GB, 16 GB
    },
    x tick label style={font=\footnotesize},
    ytick={0,25,50,75,100,125,150,175,200},
    y tick label style={font=\footnotesize},
    xlabel={Message size},
    ylabel={Bus bandwidth (GB/s)},
    label style={font=\fontsize{10}{10}\selectfont},
    grid=major,
    every major grid/.style={dashed, draw opacity=0.7},
    tick align=outside,
    legend style={
      font=\fontsize{9}{10}\selectfont,
      at={(0.99,0.02)},
      anchor=south east,
      fill=white,
      fill opacity=0.95,
      draw=gray!55,
      line width=0.4pt,
      text opacity=1,
      align=left,
      reverse legend,
    },
    scaled ticks=false,
    clip=false,
  },
  nccl axis 16M/.style={
    nccl axis, 
    xmin=15000000, 
    xtick={
      16777216, 67108864, 268435456, 1073741824, 4.294967296e9, 1.7179869184e10
    },
    xticklabels={
      16 MB, 64 MB, 256 MB, 1 GB, 4 GB, 16 GB
    },
    ymax=100, 
    ytick={0,25,50,75,100}, 
  }
}

\definecolor{nvidia_green}{RGB}{118,185,0}
\definecolor{dark_green}{RGB}{60,120,0}
\definecolor{light_green}{RGB}{170,215,80}
\definecolor{dark_gray}{RGB}{94,94,94}
\definecolor{medium_gray}{RGB}{140,140,140}
\definecolor{light_gray}{RGB}{205,205,205}
\definecolor{dark_blue}{RGB}{0, 84, 159}
\definecolor{light_blue}{RGB}{100, 170, 225}
\definecolor{cpu_blue}{RGB}{0,113,197}
\definecolor{emerald}{RGB}{0,133,100}
\definecolor{amethyst}{RGB}{93,22,130}
\definecolor{fluorite}{RGB}{250,194,0}
\definecolor{garnet}{RGB}{137,12,88}

\begin{document}
\pagestyle{empty}

\date{}

\title{NVIDIA Spectrum-X: Hardware-Accelerated Multiplane Networking for Large-Scale AI Training}
\author{
{\rm Sajy Khashab},
{\rm Albert Gran Alcoz},
{\rm Alon Gal},
{\rm Jacky Romano},
{\rm Rani Abboud},\\
{\rm Yonatan Piasetzky},
{\rm Lior Maman},
{\rm Amit Nishry},
{\rm Barak Gafni},
{\rm Omer Shabtai},
{\rm Matty Kadosh},\\
{\rm Dror Goldenberg},
{\rm Gilad Shainer},
{\rm Mark Silberstein $^{*}$}
\\[1ex]
\normalsize NVIDIA 
}



\maketitle
\thispagestyle{empty}
\begingroup
\renewcommand\thefootnote{}
\footnotetext{$^{*}$ Technion - Israel Institute of Technology and NVIDIA}
\endgroup
\begin{abstract}
As distributed model training scales to span hundreds of thousands of GPUs, scale-out networks face unprecedented performance and efficiency demands. 
NVIDIA Spectrum-X Ethernet has been designed from the ground up to achieve predictable and stable network performance with high utilization and low latency. This paper presents the Spectrum-X multiplane architecture, which replaces hierarchical depth with topological parallelism, and introduces hardware-accelerated load balancing in NICs and switches as the key architectural approach to provide fast reaction to highly dynamic network conditions at the microsecond timescales that AI training workloads demand.
 We describe the motivation, design principles, evaluation methodology and performance on state-of-the-art benchmarks, as well as the lessons we learned from deploying and debugging Spectrum-X networks in large-scale systems. Our evaluation highlights production-grade AI infrastructure performance across four core dimensions: 98\% of the theoretical line rate with low jitter-free latency;  strong cross-tenant isolation for concurrent workloads; robust, capacity-proportional bisection bandwidth and 7\% latency increase for 10\% fabric link failures; and rapid reaction to host and fabric link flaps during LLM training workloads.

  
\end{abstract}

\section{Introduction}

The rapid scaling of AI models has driven the deployment of training clusters with hundreds of thousands of GPUs, making the scale-out network a first-order bottleneck for distributed training. In Ethernet-based AI clusters, RDMA over Converged Ethernet (RoCE) has emerged as the dominant transport substrate~\cite{meta_roce_lossless,stellar-sigcomm25}.

Despite well-known limitations, the current generation of AI training systems has been able to achieve high performance with RoCE through careful engineering and tuning. These systems operate under several constraints: (i) RoCE assumes near in-order delivery, (ii) it relies on coarse-grained Equal-Cost Multi-Path (ECMP) load balancing that fails to saturate the network due to the low-entropy traffic patterns of AI workloads~\cite{meta_roce_lossless}, (iii) it uses lossless Ethernet which can induce head-of-the-line blocking and congestion propagation~\cite{dcqcn,mittal2018revisiting}, causing jitter and unstable network performance, and (iv) it relies on congestion control mechanisms such as DCQCN which are difficult to tune for collective-heavy workloads, often trading modest completion-time gains for degraded PFC behavior~\cite{meta_roce_lossless}.
Nonetheless, these challenges have been mitigated sufficiently to enable AI training at current scales.
However, ad hoc RoCE tuning is reaching its limits. Emerging AI data centers executing a single AI training workload on hundreds of thousands of GPUs pose far stricter efficiency and robustness requirements than before, and expose fundamental structural limitations in existing network fabrics.

\paragraph{Fast fabric response}
Modern training workloads are dominated by synchronous collectives, which generate highly bursty and structured traffic patterns~\cite{meta_roce_lossless}. 
With 800~Gbps and higher per-GPU network bandwidth,  flows of a few to tens of megabytes complete in $O(100)$ \usec 
so even small variations in path delay due to short-lived congestion or minor load imbalances have direct impact on the completion time. Similarly, at high fabric utilization, insufficient 
 inter- and intra-job isolation  among concurrent collectives introduces jitter, affecting collective completion time significantly.
Due to tight temporal synchronization, such transient events disproportionately impact end-to-end performance, leading to underutilization of tens of thousands of GPUs.
%
%
Metrics such as high-percentile latency and worst-flow completion time, rather than averages, best capture training efficiency~\cite{optireduce-nsdi25, swarm-nsdi25}. 
To optimize them, the fabric must react \emph{extremely quickly} to bursts, intermittent congestion, and transient load imbalances.
Existing mechanisms cannot react to real-time congestion state at the microsecond timescales that collective synchronization demands.

\paragraph{Size and bandwidth scaling}
 Reaching the scale of hundreds of thousands of GPUs in a single network necessitates adding network tiers, increasing the path length. Furthermore, scaling the bisection bandwidth under current technological constraints can be achieved via link parallelism, which increases topological complexity and reduces the available switch radix. As a result, conventional approaches to scaling the network size and bandwidth amplify latency, jitter, and load imbalance, particularly under synchronized traffic patterns. 

\paragraph{High fabric resilience}
At the required network scale, intermittent faults, link flaps and persistent anomalies become steady-state operating conditions rather than exceptions~\cite{pronto-nsdi17,netbouncer-nsdi19,holmes-nsdi25, swarm-nsdi25}.
For AI clusters, this is not merely an availability concern: the network must remain bandwidth-proportional under failures and recover quickly enough that collective performance is not repeatedly perturbed. 

\paragraph{Performance tuning at scale}
The scale and dynamism of AI clusters make performance tuning and debugging inherently challenging. The interaction between routing, congestion control, topology, and workload behavior creates a high-dimensional optimization space. Rapid and precise localization of faults and anomalies is critical for maintaining performance and reducing operational overhead.

This paper presents {\bf Spectrum-X} (\projname \cite{spx_nvidia_website}), NVIDIA's {\bf Spectrum} switches and {\bf ConnectX} NIC's based AI solution architecture for large-scale Ethernet-based training clusters. At its core, \projname is a fully hardware-accelerated multi-plane network architecture with packet-granular load balancing. The key design principle is a separation of hardware-accelerated control loops by \emph{scope}, \emph{signal}, and \emph{responsibility}. At the edge, NICs implement multi-plane load balancing together with per-plane congestion control. Inside each plane, switches perform per-packet adaptive routing to react  to local imbalance and transient congestion at packet time-scales. 
%
%
Because the NICs and in-network control loops operate on different signals and at different scopes, they can be tuned independently: switch-local decisions continuously equalize path utilization within a plane, while the NIC can run a more aggressive congestion controller across planes without being destabilized by intra-plane path transients. This decomposition is central to achieving both high utilization and strong isolation.

\projname couples this transport architecture with fast fault handling and telemetry-driven operations. The multi-plane design provides fault containment and bandwidth scaling, while fast inter-plane failover absorbs transient and permanent faults with 3~ms recovery. The same architectural regularity that enables scale also aids operations: exploiting fabric symmetry together with high-frequency telemetry enables rapid localization of faults and performance anomalies. As we show in the paper, this design yields near-perfect isolation between co-executing collectives, whether from different training jobs or separate communication streams within the same job.

We comprehensively evaluate \projname on multiple 1K-GPU testbeds designed to serve as proxies for large-scale deployments, as well as in high-fidelity simulations with up to 256,000 endpoints. In addition,
SPX is deployed in production across tens of customer clusters—including some of the largest AI clusters in the world—spanning a total of over 1 million GPUs.

\projname sustains 98\% of theoretical line rate with p99 latency of just 8–9 \usec. It demonstrates near-ideal isolation among concurrent collectives and full-scale LLM training workloads. It reaches theoretical maximum rate in complex intra-plane failures scenarios, and reacts within 3 ms to the theoretical maximum rate in case of dynamic failures.  
Last, it shows negligible performance degradation when running tightly coupled collectives over massive scale under aggressive link flap rates, both in the fabric and in the end-hosts.

\paragraph{Contributions}
This paper's main contributions are:

\begin{itemize}
    \item We identify the key challenges in networking for large-scale AI training, focusing on tail performance, scale, and failure resilience. 
    
    \item We explain the principles and details of the Spectrum-X design, including its multi-plane architecture, hardware-accelerated adaptive routing and plane load balancing.

     \item We share lessons learned from deploying Spectrum-X across multiple large-scale clusters, including our performance tuning and development methodology using proxy-scale environments.
     
    \item We evaluate Spectrum-X using state-of-the-art benchmarks and production workloads, demonstrating strong performance and resilience.
\end{itemize}
\section{Motivation}
\label{sec:motivation}

The rise of extreme-scale distributed AI exposes the fundamental limitations of existing  network infrastructure in AI data centers, posing harsh capacity and performance requirements to achieve efficient training at scale as discussed next.

\subsection{Low latency and jitter}

\providebool{simBallsInBins}
\setbool{simBallsInBins}{true}

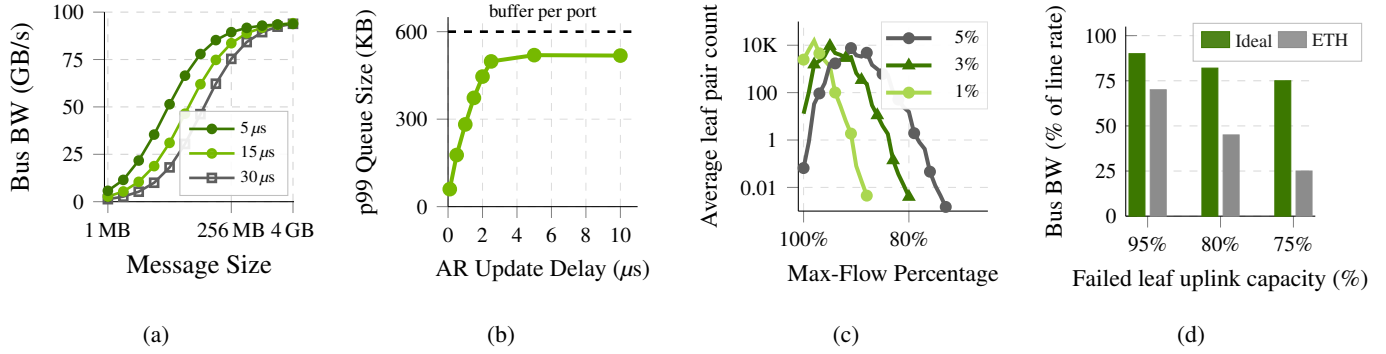
\begin{figure*}[th!]
  \centering
  \makebox[\textwidth][c]{%
    \begin{subfigure}[t]{0.23\linewidth}
      \vspace{0pt}
      \centering
      \begin{minipage}[t][4.1cm][t]{\linewidth}
        \centering
        \begin{minipage}[t]{\linewidth}
  \begin{tikzpicture}
    \begin{axis}[
      nccl axis,
      xlabel={Message Size}, 
      ylabel={Bus BW (GB/s)},
      xmode=log,
      width=\linewidth, height=\linewidth,
      xmin=900000, xmax=4.5e9,
      xtick={1048576,268435456,4294967296},
      xticklabels={1\,MB,256\,MB,4\,GB},
      ymin=0, ymax=100,
      ytick={0,25,50,75,100},
      legend pos=south east, 
      legend cell align={left},
      legend style={
        font=\scriptsize,
        fill=white, fill opacity=0.9, 
        draw=gray!50,
        nodes={scale=0.9, transform shape}
      },
    ]
      \addplot[dark_gray, line width=1pt, mark=square, mark size=1.5pt,
               mark options={solid, draw=dark_gray}] coordinates {
        (1048576,1.336)(2097152,2.672)(4194304,5.273)(8388608,10.000)
        (16777216,18.095)(33554432,30.466)(67108864,46.183)(134217728,62.235)
        (268435456,75.337)(536870912,84.198)(1073741824,89.448)
        (2147483648,92.332)(4294967296,93.843)
      };
      \addlegendentry{30\,$\mu$s}

      \addplot[nvidia_green, line width=1pt, mark=*, mark size=1.5pt,
               mark options={solid, fill=nvidia_green}] coordinates {
        (1048576,2.695)(2097152,5.389)(4194304,10.469)(8388608,18.755)
        (16777216,31.032)(33554432,46.417)(67108864,61.983)(134217728,74.792)
        (268435456,83.585)(536870912,88.889)(1073741824,91.815)
        (2147483648,93.362)(4294967296,94.156)
      };
      \addlegendentry{15\,$\mu$s}

      \addplot[dark_green, line width=1pt, mark=otimes*, mark size=1.5pt,
               mark options={solid, fill=dark_green}] coordinates {
        (1048576,5.749)(2097152,11.495)(4194304,21.712)(8388608,35.395)
        (16777216,51.404)(33554432,66.476)(67108864,77.904)(134217728,85.228)
        (268435456,89.433)(536870912,91.700)(1073741824,92.872)
        (2147483648,93.470)(4294967296,93.772)
      };
      \addlegendentry{5\,$\mu$s}
    \end{axis}
  \end{tikzpicture}
\end{minipage}
      \end{minipage}
      \subcaption{}
      \label{fig:motivation_lat}
    \end{subfigure}%
    \hfill
    \begin{subfigure}[t]{0.23\linewidth}
      \vspace{0pt}
      \centering
      \begin{minipage}[t][4.1cm][t]{\linewidth}
        \centering

  \begin{tikzpicture}
    \begin{axis}[
      nccl axis,
      width=\linewidth, height=\linewidth,
      xmin=0, xmax=11,
      xtick={0,2,4,6,8,10},
      xticklabels={0,2,4,6,8,10},
      ymin=0, ymax=650,
      ytick={0,300,600},
      xlabel={AR Update Delay ($\mu$s)},
      ylabel={p99 Queue Size (KB)},
      xlabel style={font=\small},
      ylabel style={font=\small},
      grid=major,
      grid style={line width=0.3pt, draw=gray!40},
      tick label style={font=\footnotesize},
    ]
      \addplot[nvidia_green, line width=1.5pt,
               mark=*, mark size=2pt,
               mark options={solid, fill=nvidia_green}] coordinates {
        (0.1,60) (0.5,177) (1,282) (1.5,373) (2,446) (2.5,498) (5,519.5) (10,518)
      };

      \addplot[black, dashed, line width=1pt, forget plot] coordinates {(0,600) (11,600)}
        node[pos=0.5, above, font=\scriptsize] {buffer per port};
    \end{axis}
  \end{tikzpicture}
      \end{minipage}
      \subcaption{}
      \label{fig:ar_queue_delay}
    \end{subfigure}%
    \hfill
    \begin{subfigure}[t]{0.23\linewidth}
      \vspace{0pt}
      \centering
      \begin{minipage}[t][4.1cm][t]{\linewidth}
        \centering
        \ifbool{simBallsInBins}{%

\begin{tikzpicture}
  \begin{axis}[
    nccl axis,
    width=\linewidth, height=\linewidth,
    ymode=log,
    x dir=reverse,
    xmin=65, xmax=101,
    xtick={60,80,100},
    xticklabels={60\%,80\%,100\%},
    ymin=0.001, ymax=1e5,
    ytick={0.01,1,100,10000},
    yticklabels={0.01,1,100,10K},
    ylabel={Average leaf pair count},
    xlabel={Max-Flow Percentage},
    xlabel style={font=\small},
    ylabel style={font=\small},
    grid=major,
    grid style={line width=0.3pt, draw=gray!40},
    legend style={
      font=\scriptsize,
      at={(1,1)},
      anchor=north east,
      fill=white, draw=gray!25, line width=0.3pt,
      cells={anchor=west},
      inner sep=1pt,
      row sep=0pt,
      column sep=6pt,
    },
  ]

    \addplot[light_green, line width=1.5pt,
             mark=*, mark size=1.5pt, mark repeat=3,
             mark options={solid, fill=light_green}]
      table[x=pct, y=fr_1, col sep=comma]
      {figures/ballsinbins/histogram_256l_128s_data.csv};
    \addlegendentry{1\%}


    \addplot[dark_green, line width=1.5pt,
             mark=triangle*, mark size=1.5pt, mark repeat=3,
             mark options={solid, fill=dark_green}]
      table[x=pct, y=fr_3, col sep=comma]
      {figures/ballsinbins/histogram_256l_128s_data.csv};
    \addlegendentry{3\%}


    \addplot[dark_gray, line width=1.5pt,
             mark=otimes*, mark size=1.5pt, mark repeat=3,
             mark options={solid, fill=dark_gray}]
      table[x=pct, y=fr_5, col sep=comma]
      {figures/ballsinbins/histogram_256l_128s_data.csv};
    \addlegendentry{5\%}

  \end{axis}
\end{tikzpicture}%
        }{%
          \begin{tikzpicture}
  \begin{axis}[
    width=\linewidth, height=5.5cm,
    ymode=log,
    x dir=reverse,
    xmin=69, xmax=101.5,
    xtick={70,75,80,85,90,95,100},
    xticklabels={70\%,75\%,80\%,85\%,90\%,95\%,100\%},
    xticklabel style={font=\footnotesize},
    ymin=0.8, ymax=200000,
    ytick={1,10,100,1000,10000,100000},
    yticklabels={1,10,100,1K,10K,100K},
    ylabel={Count leaf pairs},
    xlabel={Max-flow \% of full BW},
    xlabel style={font=\small},
    ylabel style={font=\small},
    grid=major,
    grid style={line width=0.3pt, draw=gray!40},
    legend style={
      font=\scriptsize,
      at={(0.97,0.97)},
      anchor=north east,
      fill=white, draw=gray!50,
      cells={anchor=west},
    },
  ]
    \addplot[nvidia_green, line width=1.5pt, mark=*, mark size=1.5pt,
             mark options={solid, fill=nvidia_green}] coordinates {
      (100,21898)(96.875,9168)(93.75,4791)(90.625,1550)
      (87.5,424)(84.375,147)(81.25,31)(78.125,11)(75,2)
    };
    \addlegendentry{PHX61 2024-7-12}
    \addplot[nvidia_green, draw=none, mark=none, forget plot,
             nodes near coords, nodes near coords align={anchor=south},
             every node near coord/.append style={font=\scriptsize, color=nvidia_green!80!black, inner sep=3pt, xshift=4pt},
             point meta=explicit symbolic,
    ] coordinates { (100,21898) [21898] };
    \addplot[nvidia_green, draw=none, mark=none, forget plot,
             nodes near coords, nodes near coords align={anchor=north},
             every node near coord/.append style={font=\scriptsize, color=nvidia_green!80!black, inner sep=3pt, xshift=-4pt},
             point meta=explicit symbolic,
    ] coordinates {
      (90.625,1550) [1550]
      (81.25,31)    [31]
    };
    \addplot[nvidia_green, draw=none, mark=none, forget plot,
             nodes near coords, nodes near coords align={anchor=north},
             every node near coord/.append style={font=\scriptsize, color=nvidia_green!80!black, inner sep=3pt, xshift=-4pt, yshift=2pt},
             point meta=explicit symbolic,
    ] coordinates { (75,2) [2] };

    \addplot[nvidia_green!50!black, line width=1.5pt, mark=square*, mark size=1.5pt,
             mark options={solid, fill=nvidia_green!50!black}] coordinates {
      (100,6206)(96.875,6462)(93.75,4199)(90.625,2005)
      (87.5,1037)(84.375,390)(81.25,129)(78.125,16)(75,3)(71.875,1)
    };
    \addlegendentry{PHX11 2024-8-7}
    \addplot[nvidia_green!50!black, draw=none, mark=none, forget plot,
             nodes near coords, nodes near coords align={anchor=north},
             every node near coord/.append style={font=\scriptsize, color=nvidia_green!50!black, inner sep=3pt, xshift=4pt},
             point meta=explicit symbolic,
    ] coordinates { (100,6206) [6206] };
    \addplot[nvidia_green!50!black, draw=none, mark=none, forget plot,
             nodes near coords, nodes near coords align={anchor=south},
             every node near coord/.append style={font=\scriptsize, color=nvidia_green!50!black, inner sep=3pt, xshift=4pt},
             point meta=explicit symbolic,
    ] coordinates {
      (90.625,2005) [2005]
      (81.25,129)   [129]
      (75,3)        [3]
    };
  \end{axis}
\end{tikzpicture}%
        }
      \end{minipage}
      \subcaption{}
      \label{subfig:ballsinbins}
    \end{subfigure}%
    \hfill
    \begin{subfigure}[t]{0.23\linewidth}
      \vspace{0pt}
      \centering
      \begin{minipage}[t][4.1cm][t]{\linewidth}
        \centering
        \begin{tikzpicture}
  \begin{axis}[
    ybar,
    width=\linewidth,
    height=\linewidth,
    bar width=6pt,
    ymin=0, ymax=105,
    ytick={0, 25, 50, 75, 100},
    ylabel={Bus BW (\% of line rate)},
    symbolic x coords={95pct, 80pct, 75pct},
    xtick={95pct, 80pct, 75pct},
    xticklabels={95\%, 80\%, 75\%},
    x tick label style={font=\footnotesize},
    y tick label style={font=\footnotesize},
    xlabel={Failed leaf uplink capacity (\%)},
    axis x line*=bottom,
    axis y line*=left,
    ymajorgrids=true,
    grid style={dashed, gray!30},
    enlarge x limits=0.15,
    xlabel style={font=\small},
    ylabel style={font=\small, align=center},
    legend image code/.code={\fill[#1] (0,-0.07cm) rectangle (0.08cm,0.14cm);},
    legend style={
      font=\scriptsize,
      at={(0.8,1.02)},        
      anchor=north,
      fill=white,
      fill opacity=0.9,
      draw=gray!50,
      row sep=2pt,
    },
    legend columns=2,
  ]

  \addplot[fill=dark_green, draw=dark_green, bar shift=-4pt] coordinates {
    (95pct,90) (80pct,82) (75pct,75)
  };
  \addlegendentry{Ideal}

  \addplot[fill=medium_gray, draw=medium_gray, bar shift=4pt] coordinates {
    (95pct,70) (80pct,45) (75pct,25)
  };
  \addlegendentry{ETH}

  \end{axis}
\end{tikzpicture}
      \end{minipage}
      \subcaption{}
      \label{subfig:perf_under_failures}
    \end{subfigure}%
  }

  \caption{
    (\ref{fig:motivation_lat}) Impact of network latency on \ATOA collective (256-endpoint simulation). 
    (\ref{fig:ar_queue_delay}) Impact of switch load balancing delay on queue size.
    (\ref{subfig:ballsinbins}) Leaf-to-leaf max-flow distribution simulation.
    (\ref{subfig:perf_under_failures}) \ATOA bandwidth under partial uplink failure.
  }
  \label{fig:motivation_network}
\end{figure*}
It is well known that the Collective Completion Time (CCT) of synchronous collectives (\AR, \AG, \ATOA) is determined by network stragglers~\cite{lin2025understandingstragglerslargemodel} - individual slow flows that delay the CCT, and in turn collective performance affects the performance of the entire training run. 

As network bandwidth scales beyond 800~Gbps, we observe that collectives are also becoming highly sensitive to network \emph{latency}. 
\Cref{fig:motivation_lat} illustrates the impact of network latency on \ATOA collectives. Higher latencies significantly degrade the collective performance for message sizes commonly used in training collectives \cite{meta_roce_lossless,alibaba_hpn}.
Similar behavior is observed for Ring All-Reduce and other collectives.

Latency jitter, i.e., latency variations over time, is another significant performance factor. Jitter is a result of poor fabric performance isolation and load balancing. Intra-collective jitter arises due to congestion within a single collective, whereas inter-collective jitter occurs in a loaded network when neighboring traffic degrades victim throughput. Finally, jitter is a result of variable queuing depth as the network alternates between underloaded and overloaded states.

We provide concrete production examples of such jitter and its diagnosis via bandwidth histograms and high-frequency telemetry in \Cref{ssec:histograms_stragglers} and \Cref{ssec:lessons_hft}.

Maintaining jitter-free low latency in AI training workloads is hard: collectives generate coordinated traffic bursts that arrive simultaneously at switch ports. Thus, queue build-up and latency increase are inevitable, unless \emph{the network fabric is able to quickly balance the load at sub-RTT timescales}.

The graph in \Cref{fig:ar_queue_delay} shows the impact of load balancing delay on the  queue size in a switch that implements per-packet Join-Shortest-Queue load balancing  among its 256 egress ports. For 100~ns delay the queues are small, however even at 1\usec they grow almost 5-fold, and saturate at 2.5\usec because at that point the decisions are random in practice. Such queues add about 20\usec to the tail latency, and significantly impact the collective performance (\Cref{fig:motivation_lat}).

\subsection{Network scale}
\label{subs:scale}

Multiplane topologies~\cite{alibaba_hpn,deepseek-v3}
address the challenge to scale networks to hundreds of thousands of nodes by decomposing each NIC bandwidth into multiple lower-speed interfaces (e.g., 4$\times$200 Gbps from a single 800 Gbps ConnectX-8 NIC), each connecting into an independent \emph{network plane}.
Each plane is typically realized as a two-tier fat tree, but can also be extended to a multi-pod three-tier fat-tree as well.
At this scale, the maximal size of a two-tier topology is increased to 128K, and the respective three-tier to 16M.

Multiplane topology, however, is only effective if the traffic is perfectly balanced across the planes. Achieving such a balance is challenging. 
One common approach is to partition collective operations across planes at the software level~\cite{nccl_docs}, assigning different message transfers to different planes through explicit scheduling in the collective library.
Such approaches operate at RDMA message granularity, which is too coarse for optimal balance, and require software modifications.

Using per-packet spraying among the planes below the transport layer~\cite{uec_uet_spec,Adaptive_routing} achieves better scheduling granularity. However, it uses a single transport CC loop, and assigns packets to planes without visibility into the per-plane congestion state. When a plane is degraded, i.e., due to fabric link flaps, the load symmetry among the planes is violated. Thus the performance of the load-oblivious spraying is dictated by the slowest plane. Worse, this approach dramatically increases the blast radius of the in-fabric link failure, which is a common place at such scales. 

More critically,
a transient plane failure due to a link flap at the host leads to a stall of the traffic from all the planes, since a large portion of the packets from all the flows get dropped.
Host-based non-HW-accelerated approaches, have slow reaction to plane failure, and therefore their slow recovery incurs high performance loss as shown in Section~\ref{ssec:large_scale_Resiliency}.

Inter-plane load balancing must therefore react quickly and precisely to per-plane congestion and transient failures.

\subsection{Time-to-AI}
\label{ssec:time-to-ai}

Minimizing \emph{Time-to-AI}, i.e., the time from the moment GPUs are installed at the facility until they run a full-scale training job, is a crucial goal in modern AI training clusters.
Unlike conventional hyperscale data centers that afford gradual growth over 1–2 years to reach capacity\cite{zhao-nsdi19-rewiring}
the timescales for building modern-scale AI factories are over \emph{an order of magnitude} shorter.  For example, X-AI Colossus has achieved its full capacity in four months~\cite{xai_colossus,hpcwire_colossus}.

Network fabric is perhaps the most challenging part to get up and running quickly, as it requires connecting hundreds of thousands of fragile optical cables. A non-negligible portion of these cables malfunction. For example, dust at the construction site might affect optical connectors, and manifest itself as link flaps, inability to establish connectivity, or high Bit Error Rate (BER).
Such faulty links are proactively disabled to prevent them from degrading cluster performance.

Link failures are inevitable, but their effect on the cluster performance should be proportional to the reduction in the physical connectivity. This is challenging, as failed links in the fabric may result in multiple source–destination pairs experiencing significant connectivity degradation, due to the combined impact of faults along the path.

\Cref{subfig:ballsinbins} shows the maximum flow between leaf-pairs using a simulation of a 32K endpoint leaf-spine topology for different percentages of uniformly distributed random link failures.
%
%
%
%
This degradation in capacity introduces extreme bandwidth asymmetry in the network that is especially present during early stages of cluster life and strongly affects the Time-to-AI.
\Cref{subfig:perf_under_failures} measures the performance impact of such connectivity loss in an \ATOA collective running over a 480-endpoint H100 cluster.
We systematically drop uplinks from a single leaf switch, varying the fraction of remaining uplink capacity (x-axis), and measure the achieved collective bandwidth.
Traditional Ethernet solutions degrade in a non-proportional way to the bandwidth loss, particularly when asymmetry is severe, as is common in early-stage clusters.

\section{\projname Overview}
\label{sec:design}

\begin{figure}[htbp]
\centering
\includegraphics[width=\linewidth]{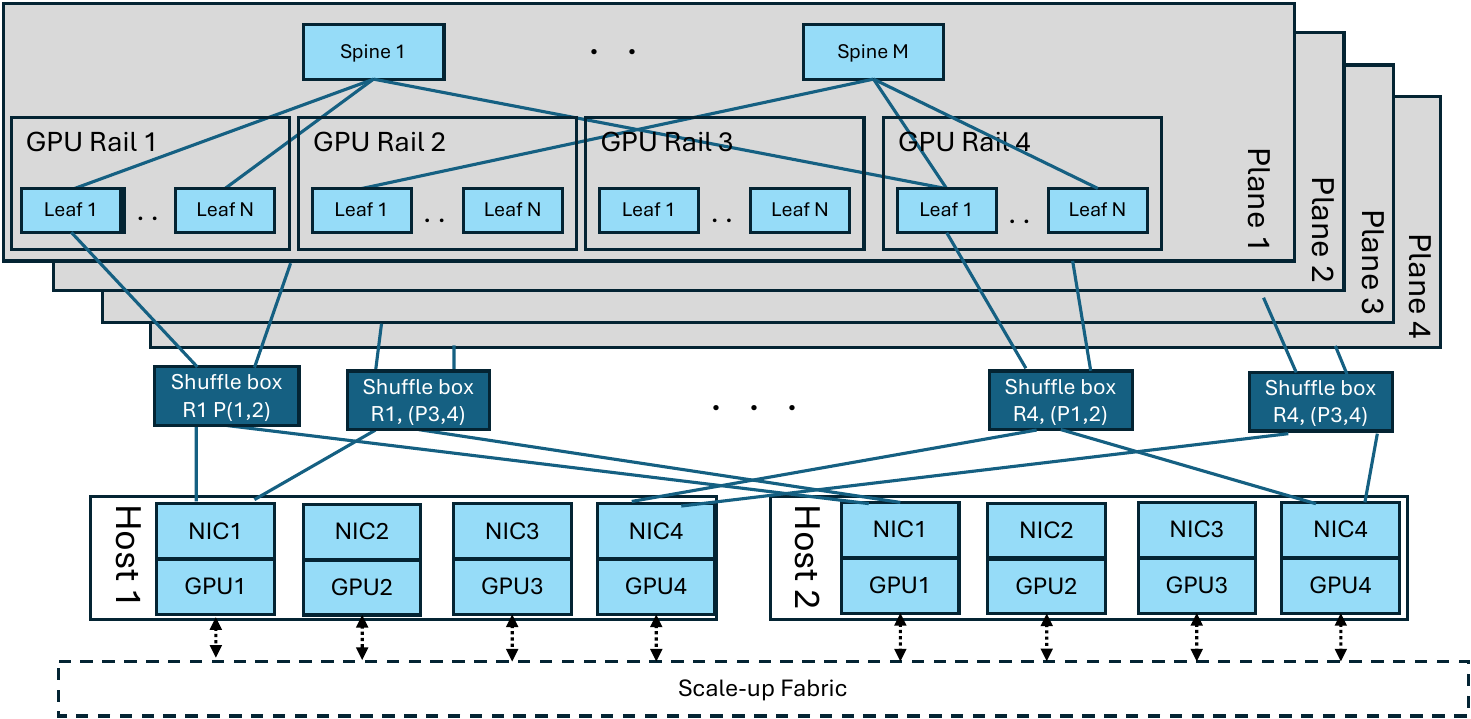}
\caption{\projname topology overview: 2-level FT Multiplane rail-optimized topology for 4 disconnected planes. Larger, 3-level topologies keep rail separation in the spines, connecting on the 3rd tier.}
\label{fig:spx-overview}
\end{figure}


    

Spectrum-X (SPX) is NVIDIA's reference architecture for scale-out GPU networks. 
It started from the Hopper generation (H100)~\cite{spx_h100_announcement} and evolved alongside next generations of Grace-Blackwell (GB200), Grace-Blackwell Ultra (GB300), and lately Vera-Rubin (VR200).
This section focuses on the Spectrum-X architecture for the Blackwell Ultra generation. 
The discussion of the north-south ~\cite{blog_ns}, scale across ~\cite{blog_xgs}, scale-up (NVLINK), storage ~\cite{blog_spx_storage}, and computing technologies is out of scope.

\projname implements hardware-accelerated load balancing within the network via switch Adaptive Routing (AR), and across planes via the NIC Plane Load Balancer (PLB). Both mechanisms utilize fine-grained per-packet load balancing.
Congestion in the network is controlled with combined lossless (PFC) and sender-based congestion control rate limiting mechanism, co-designed to preserve low tail latency at high load and present robust, linear performance with respect to available bisection bandwith under failures.

Packet-level load balancing schemes introduce out-of-order packet arrival~\cite{mcclure2025load}, which the transport in \projname handles via direct data placement without buffering data in the receiver NIC and by reordering completions transparently to applications.


The non-blocking topology and Spectrum-X load-balancing features are designed to remain \emph{job-allocation agnostic} inside the data center. They aim to provide consistent performance regardless of the physical alignment of the workloads to the collective patterns and locality, while ensuring performance isolation for neighboring jobs, simplifying the job-scheduling restrictions, and improving cluster utilization.  

\subsection{Network topology}
\label{sec:design:topo}
Spectrum-X topology is based on a multi-plane design (\cref{fig:spx-overview}), where planes are interconnected at the host side (e.g., via a shuffle-box), enabling high-BW fabrics with uniform link speeds.  

\paragraph{Single plane network}
We utilize leaf-spine or fat-tree topologies.
Topologies are typically non-blocking and rail-optimized.
We utilize \emph{parallel links} between switches when building non-max-scale topologies to improve port utilization and achieve better resiliency to link failures.

\paragraph{Plane separation}
In the \projname topology planes are disconnected, with each NIC connected to all planes, utilizing passive optical shuffle-boxes to reduce cabling.
This design preserves full NIC-to-NIC reachability while exposing substantially more path diversity at the host edge. 
Plane separation is the key enabler for scaling a two-tier fabric to cluster sizes that typically required  an additional network tier.



The multi-plane design requires a new load-balancing layer at the NIC. 
Load balancing decisions now span not only paths inside a plane, but also the choice of the plane itself.
This NIC-level balancing mechanism is a first-order design consideration as discussed below.

\subsection{Design principles}
\label{sec:designprinciples}




\paragraph{Hardware acceleration of per-packet load balancing}
As link rates reach 800\,Gbps and beyond, bandwidth-delay products grow while propagation delay -- governed by the speed of light -- remains fixed. Effective congestion avoidance requires reacting before in-flight data exceeds buffer capacity at the congestion point.

Spectrum-X targets three such congestion points with their respective congestion avoidance control loops:
 \begin{itemize}[leftmargin=*,noitemsep,nosep]
\item \textbf{Switch: fabric ports}, where per-packet adaptive routing prevents in-switch queue buildup;
\item \textbf{Switch: endpoint ports}, where congestion control throttles senders before egress buffers saturate; and
\item \textbf{NIC: plane ports}, where plane load balancing avoids cross-plane imbalance per packet.
\end{itemize}
Closing all three control loops  within their BDP budgets is beyond the reach of software control paths. Therefore, hardware acceleration is a structural requirement of the fine-grain load balancing architecture, not a performance optimization.

\paragraph{Separate Problems, Separate Control Loops}
Handling the three control loops simultaneously creates complex interplay between their respective controllers.
A micro-burst at a fabric port raises the RTT that congestion control measures; this can be misread as endpoint incast and throttle the sender across all planes, or erroneously trigger plane load balancing to migrate traffic away from an otherwise healthy plane.

We address this  by decomposing control into separate loops with distinct objectives, signals, and timescales:
\begin{itemize}[leftmargin=*,noitemsep,nosep]
\item \textbf{Adaptive Routing:} stateless, per-packet load balancing reacting within hundreds of nanoseconds.
\item \textbf{Congestion Control:} stateful, per-destination rate control at RTT timescales, triggered only by incast that cannot be resolved in-network.
\item \textbf{Plane Load Balancing:} per-packet NIC-based balancing across planes, combining per-plane congestion signals with local queue feedback.
\end{itemize}

This separation of concerns avoids interference between the load balancing and congestion control loops and allows each to operate at its natural control horizon.


\paragraph{Appropriate state granularity for fast convergence}
%
Fast fabric convergence, i.e., quickly determining the forwarding path and the sending rate given  network conditions, 
is essential for high end-to-end performance.

\projname adopts stateless, flow-agnostic intra-plane load balancing and stateful, per-destination inter-plane load balancing. The former reacts at sub-RTT timescales to local imbalance without host involvement, converging quickly to a packet-granular solution within the plane; the latter is oblivious to the intra-plane paths, it leverages NIC acceleration and per-plane congestion control to adapt across a small number of planes within a few RTTs.

This design maintains minimal but sufficient state at the NIC: per-destination, per-plane congestion control. A single global controller would fail to adapt to inter-plane asymmetry.

\paragraph{Hardware-accelerated fault-tolerance}
At massive scales, there is always some fraction of degraded links, some disabled permanently, some experiencing a transient disruption (a flap). 

\projname strives to react as quickly as possible to \emph{transient failures} and converge to the degraded capacity-proportional bandwidth, whereas the handling of \emph{permanent failures} is pushed off the critical path.
Therefore, handling transient failures is hardware-accelerated: adaptive routing excludes a locally failed fabric link within O(100)\,ns, and the plane load balancer stops sending on an inaccessible plane within a few RTTs. 
Permanent failures are handled at software timescales, and involve
computing forwarding weights proportional to remaining healthy capacity (\Cref{sec:design:fault_tolerance}).



\paragraph{Handling asymmetries: stateless fabric, stateful planes}
Failures introduce asymmetries across the network. Asymmetries between paths within the plane and between planes are inherently different. Intra-plane asymmetries are typically small due to high radix, while inter-plane asymmetries are larger due to the limited number of planes.
For example, a failure of one leaf uplink out of a few hundred increases load on remaining uplinks by $<1\%$, while a failure of a single plane out of four overloads the remaining planes disproportionately.

Whether a failure requires traffic redistribution depends on load: lightly loaded planes can absorb capacity loss, while heavily loaded ones cannot. Plane load balancing must therefore account for both topology and traffic.

Traffic composition introduces further asymmetry: a small fraction of control-plane traffic requires in-order delivery and cannot use per-packet load balancing. However, production telemetry shows that over 97\% of traffic tolerates out-of-order delivery, so this constraint has minimal impact.
These flows need no hardware acceleration, and they can be rate-limited, so the standard host network stack suffices.


Accordingly, we use stateless adaptive routing within the fabric to ensure scalability and avoid large per-path state within switches, and stateful per-plane control at the NIC to handle inter-plane imbalance efficiently.

\paragraph{Lossless Link Layer}
A lossless fabric is preferred because per-packet load balancing introduces out-of-order delivery, making packet loss indistinguishable from reordering. Enforcing lossless operation removes this ambiguity: in a healthy network, losses are rare and typically confined to failure scenarios, thereby simplifying recovery.

PFC storms~\cite{mittal2018revisiting} are often cited as a common argument against lossless fabrics.
We observe no evidence of such storms in our own operational experience across more than two years of production operation of \projname deployed on the largest AI clusters.
Similar conclusions have been reported in other large-scale deployments~\cite{meta_roce_lossless,microsoft_roce_lossless}.
As a safeguard against misconfigurations and misbehaving devices, we deploy a PFC watchdog~\cite{sonic_pfc_watchdog} that detects and mitigates PFC storms.
Congestion caused by slow receivers, which can cause PFC storms, is handled by congestion control, which throttles senders before the congestion spreads through the network (\cref{subsection:design:cc}).

\section{Forwarding and load balancing}


\subsection{Per-packet Adaptive Routing}

\projname switches implement per-packet Adaptive Routing (AR) via a quantized approximation of Join-Shortest-Queue (JSQ)~\cite{Adaptive_routing,drill-sigcomm17}: for each arriving packet, the switch scores every egress port in the ECMP group by its current queue depth, sampled at sub-microsecond intervals, and forwards the packet to one of the least-congested ports 

We further extend this mechanism with \emph{weighted-adaptive load balancing} ~\cite{wcmp-eurosys14,juniper_ai_dc_wp}  to account for remote capacity imbalance to a given destination caused by remote failures of links in the fabric, as described in \cref{sec:design:fault_tolerance}.

\subsection{Congestion Control}
\label{subsection:design:cc}
\projname congestion control is tailored for AI collectives, prioritizing short completion time over steady-state or extreme-incast convergence.
It handles three types of congestion: incast, fabric congestion, and slow receivers.

\paragraph{Incast}
Collective traffic is short, synchronized, and bursty: bandwidth-optimal collectives generate synchronized bursts across a few high-rate
flows rather than sustained incast.
Conventional CC algorithms (e.g., DCQCN~\cite{dcqcn}, Swift~\cite{swift_google}) either overreact to short synchronized collective
bursts, throttling senders and failing to recover in a timely fashion within a single collective, or fail to react, leading to packet drops.

\projname CC handles the resulting endpoint congestion with RTT probes and sender-side rate control: the sender maintains a dynamic RTT target
that guides its rate increases and decreases; the exact update rules are beyond the scope of this paper.

\begin{figure}[htbp]
    \centering
    \includegraphics[width=1\linewidth]{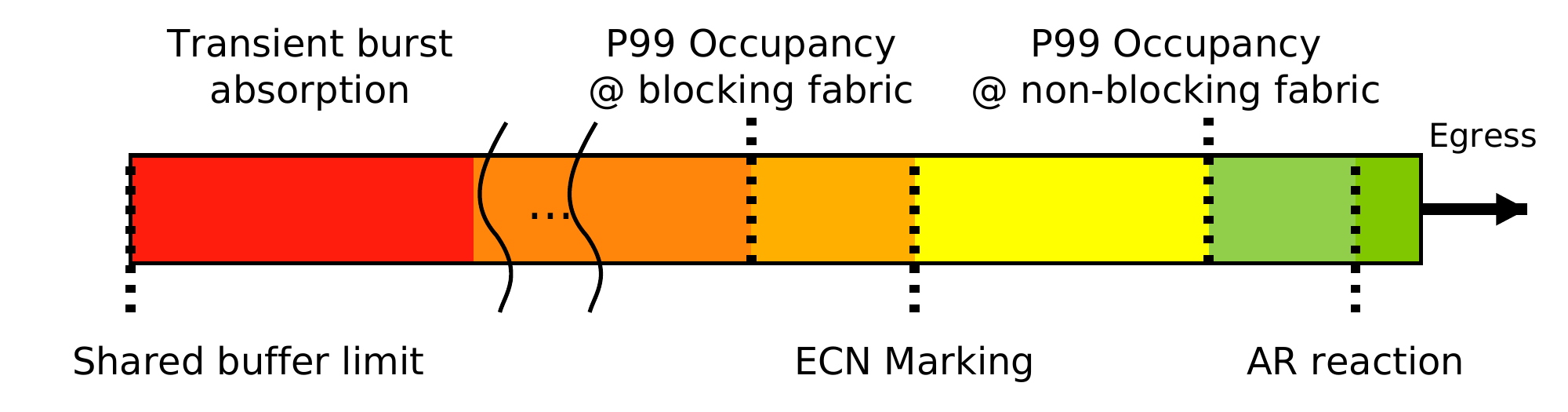}
    \caption{Congestion signaling and load balancing reaction points in switches (growing from right to left).}
    \label{fig:switch_congestion_trshold}
\end{figure}

\paragraph{Fabric congestion}
\projname topologies are non-blocking by design.
While the fabric remains non-blocking, AR balances load across paths.
However, inevitable fabric failures can make the network blocking; queues then grow beyond the level expected from AR alone, and CC must
reduce the sending rate.
To achieve this behavior, \projname CC combines RTT probes with explicit switch signals: ECN thresholds are configured to afford enough time
for the AR logic to react to short-lived load imbalance.
Thus, ECN marks only when load-balancing capacity is exhausted (\cref{fig:switch_congestion_trshold}), and CC reacts to fabric congestion
only upon the resulting Congestion Notification Packets (CNPs).
\Cref{sec:design:fault_tolerance} describes how \projname handles fabric failures.

\paragraph{Slow receivers}
Backpressure from the GPU or the intra-server interconnect can slow a receiver, building queues first in the NIC receive buffers and then in
the network.
RTT probes capture this queuing, so CC throttles senders to the slow receiver before PFC pauses propagate through the network and affect
other flows.
Switch and NIC buffer headroom is provisioned to give CC enough time to react before the PFC watchdog engages and falsely flags such cases as
PFC storms.
Persistent slow receivers generally result from misconfiguration or poor application behavior and must be fixed at their source.

\subsection{Plane Load Balancing}
\label{sec:design:PLB}

\paragraph{Application Model}
Multiplane is transparent to applications.
Each NIC exposes a single RoCE device regardless of the number of underlying planes.
Applications open queue pairs (QPs) on that device using standard RDMA verbs;
all traffic distribution, load balancing, and plane resiliency are handled entirely in NIC hardware.
Collective communication libraries such as NCCL and the transport layer require no modification.

\paragraph{Plane selection}
Plane load balancing combines adaptive routing with end-to-end congestion control.

For each destination, the NIC maintains four independent CC contexts, one per plane, shared by all QPs to that destination.
Each context independently issues RTT probes on its assigned plane and processes incoming CNPs to calculate its plane's rate allowance.
This per-destination per-plane state lets the NIC distinguish congestion on one plane from healthy conditions on another, rather than reacting uniformly across all planes.

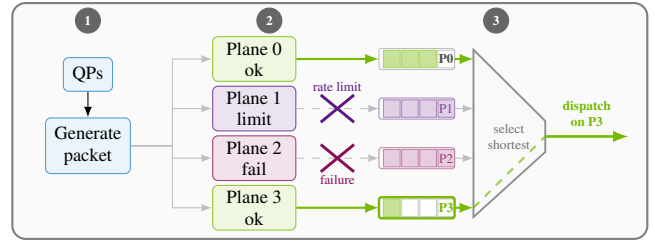
\begin{figure}[t]
\centering
\begin{tikzpicture}[
  yscale=0.82,
  nicbnd/.style  = {draw=black!50, rounded corners=5pt,
                    fill=gray!4, line width=0.8pt},
  stepnum/.style = {draw=black!60, fill=black!60, circle, text=white,
                    font=\bfseries\tiny, minimum size=0.33cm, inner sep=0pt},
  ccok/.style    = {draw=nvidia_green!80, fill=nvidia_green!12, rounded corners=2pt,
                    minimum width=1.10cm, minimum height=0.49cm,
                    align=center, font=\scriptsize, inner sep=2pt},
  cclimit/.style = {draw=amethyst!80, fill=amethyst!12, rounded corners=2pt,
                    minimum width=1.10cm, minimum height=0.49cm,
                    align=center, font=\scriptsize, inner sep=2pt},
  ccfail/.style  = {draw=garnet!80, fill=garnet!12, rounded corners=2pt,
                    minimum width=1.10cm, minimum height=0.49cm,
                    align=center, font=\scriptsize, inner sep=2pt},
  qslot/.style   = {draw=gray!40, fill=white,
                    minimum width=0.22cm, minimum height=0.23cm, inner sep=0pt},
  qfull/.style   = {draw=nvidia_green!60, fill=nvidia_green!40,
                    minimum width=0.22cm, minimum height=0.23cm, inner sep=0pt},
  qlimit/.style  = {draw=amethyst!50, fill=amethyst!20,
                    minimum width=0.22cm, minimum height=0.23cm, inner sep=0pt},
  qfail/.style   = {draw=garnet!50, fill=garnet!20,
                    minimum width=0.22cm, minimum height=0.23cm, inner sep=0pt},
  arr/.style     = {->, >=latex, thin},
  okarr/.style   = {->, >=latex, line width=0.8pt, color=nvidia_green},
  garr/.style    = {->, >=latex, thin, color=gray!50},
  slbl/.style    = {font=\tiny, align=center, text=gray!60}
]

\node[nicbnd, minimum width=8.4cm, minimum height=3.12cm]
  at (4.30, 1.10) {};

\node[stepnum] at (1.10, 2.72) {1};
\node[stepnum] at (3.50, 2.72) {2};
\node[stepnum] at (6.50, 2.72) {3};

\node[draw=cpu_blue!70, fill=cpu_blue!8, rounded corners=2pt,
      minimum width=0.65cm, minimum height=0.33cm,
      align=center, font=\scriptsize] (qp) at (1.10, 1.85) {QPs};
\node[draw=cpu_blue!70, fill=cpu_blue!8, rounded corners=2pt,
      minimum width=1.10cm, minimum height=0.37cm,
      align=center, font=\scriptsize] (pktgen) at (1.10, 0.70)
        {Generate\\packet};
\draw[arr] (qp.south) -- (pktgen.north);

\node[ccok]    (cc0) at (3.30, 2.10) {Plane 0\\ok};
\node[cclimit] (cc1) at (3.30, 1.30) {Plane 1\\limit};
\node[ccfail]  (cc2) at (3.30, 0.50) {Plane 2\\fail};
\node[ccok]    (cc3) at (3.30, -0.30) {Plane 3\\ok};

\draw[garr] (pktgen.east) -- ++(0.55,0) |- (cc0.west);
\draw[garr] (pktgen.east) -- ++(0.55,0) |- (cc1.west);
\draw[garr] (pktgen.east) -- ++(0.55,0) |- (cc2.west);
\draw[garr] (pktgen.east) -- ++(0.55,0) |- (cc3.west);

\draw[okarr] (cc0.east) -- (4.95, 2.10);
\draw[okarr] (cc3.east) -- (4.95, -0.30);

\draw[garr, dashed] (cc1.east) -- (4.95, 1.30);
\draw[garr, dashed] (cc2.east) -- (4.95, 0.50);

\draw[color=amethyst, line width=1.1pt] (4.18, 1.08) -- (4.62, 1.52);
\draw[color=amethyst, line width=1.1pt] (4.18, 1.52) -- (4.62, 1.08);
\node[slbl, text=amethyst] at (4.40, 1.65) {rate limit};

\draw[color=garnet, line width=1.1pt] (4.18, 0.28) -- (4.62, 0.72);
\draw[color=garnet, line width=1.1pt] (4.18, 0.72) -- (4.62, 0.28);
\node[slbl, text=garnet] at (4.40, 0.16) {failure};


\node[qfull]  at (5.12, 2.10) {};
\node[qfull]  at (5.36, 2.10) {};
\node[qfull]  at (5.60, 2.10) {};
\node[qslot]  at (5.84, 2.10) {};
\draw[draw=gray!50, rounded corners=1pt]
  (4.95, 1.93) rectangle (5.95, 2.27);
\node[font=\tiny\bfseries, text=black!70] at (5.84, 2.10) {P0};

\node[qlimit]  at (5.12, 1.30) {};
\node[qlimit]  at (5.36, 1.30) {};
\node[qlimit]  at (5.60, 1.30) {};
\node[qlimit]  at (5.84, 1.30) {};
\draw[draw=amethyst!50, rounded corners=1pt]
  (4.95, 1.13) rectangle (5.95, 1.47);
\node[font=\tiny, text=amethyst!80] at (5.84, 1.30) {P1};

\node[qfail]  at (5.12, 0.50) {};
\node[qfail]  at (5.36, 0.50) {};
\node[qfail]  at (5.60, 0.50) {};
\node[qfail]  at (5.84, 0.50) {};
\draw[draw=garnet!50, rounded corners=1pt]
  (4.95, 0.33) rectangle (5.95, 0.67);
\node[font=\tiny, text=garnet!80] at (5.84, 0.50) {P2};

\node[qfull]  at (5.12, -0.30) {};
\node[qslot]  at (5.36, -0.30) {};
\node[qslot]  at (5.60, -0.30) {};
\node[qslot]  at (5.84, -0.30) {};
\draw[draw=nvidia_green, line width=0.9pt, rounded corners=1pt]
  (4.95, -0.47) rectangle (5.95, -0.13);
\node[font=\tiny\bfseries, text=nvidia_green] at (5.84, -0.30) {P3};

\draw[okarr] (5.95, 2.10) -- (6.20, 2.10);
\draw[garr]  (5.95, 1.30) -- (6.20, 1.30);
\draw[garr]  (5.95, 0.50) -- (6.20, 0.50);
\draw[okarr] (5.95, -0.30) -- (6.20, -0.30);

\draw[draw=black!45, fill=gray!5, line width=0.7pt]
  (6.20, 2.25) -- (6.20, -0.45) -- (7.15, 0.60) -- (7.15, 1.10) -- cycle;

\draw[nvidia_green, line width=0.8pt, dashed, opacity=0.7]
  (6.20, -0.30) -- (7.15, 0.85);

\node[font=\tiny, text=black!55, align=center, text width=0.8cm]
  at (6.67, 0.85) {select shortest};

\draw[okarr, line width=1.0pt]
  (7.15, 0.85) -- (8.25, 0.85)
  node[midway, above, font=\tiny\bfseries, color=nvidia_green, align=center]
    {dispatch\\on P3};

\end{tikzpicture}
\caption{NIC per-packet plane selection:
\textbf{(1)}~generate packet from QPs,
\textbf{(2)}~query CC contexts and mask rate-limited or failed planes,
\textbf{(3)}~choose shallowest eligible egress queue.}
\label{fig:plb-selection}
\end{figure}

When a packet is ready for transmission, the NIC performs a two-stage hierarchical plane selection (\cref{fig:plb-selection}).
\textbf{Rate filter (E2E congestion):} The CC rate allowance for each plane is compared against the current transmission rate (computed based on transmission time of last packet burst). Planes whose allowance falls below that rate are excluded, yielding an eligible set of uncongested planes.
\textbf{Local queue selection:} From the eligible planes, the NIC selects the plane with shallowest local egress queue, mirroring switch adaptive routing.

This hierarchy ensures that E2E congestion state takes precedence: congested planes are excluded before local queue depth is consulted. Thus, congestion on one plane lowers only that plane's allowance for the destination and diverts its traffic to the other planes, whereas endpoint incast congests all four contexts and throttles all traffic to that destination.


\subsection{Resiliency}
\label{sec:design:fault_tolerance}


\subsubsection{Endpoint failure and recovery}
Recovery from host plane failures must be fast to avoid CCT degradation. Transient link flaps that resolve within O(ms) should be absorbed transparently by the network.
Local host plane failures are detected directly by the NIC PLB, enabling immediate in-hardware reaction.
Remote host plane failures are detected via consecutive RTT probe timeouts on that plane. Once a threshold is reached, the plane is removed from the eligible set and no new traffic is forwarded over it.
Detection is fully handled in NIC hardware and completes within a few RTTs, prior to control-plane involvement. 

\subsubsection{Fabric link failures}
\begin{figure}
    \centering
    \includegraphics[width=0.70\linewidth]{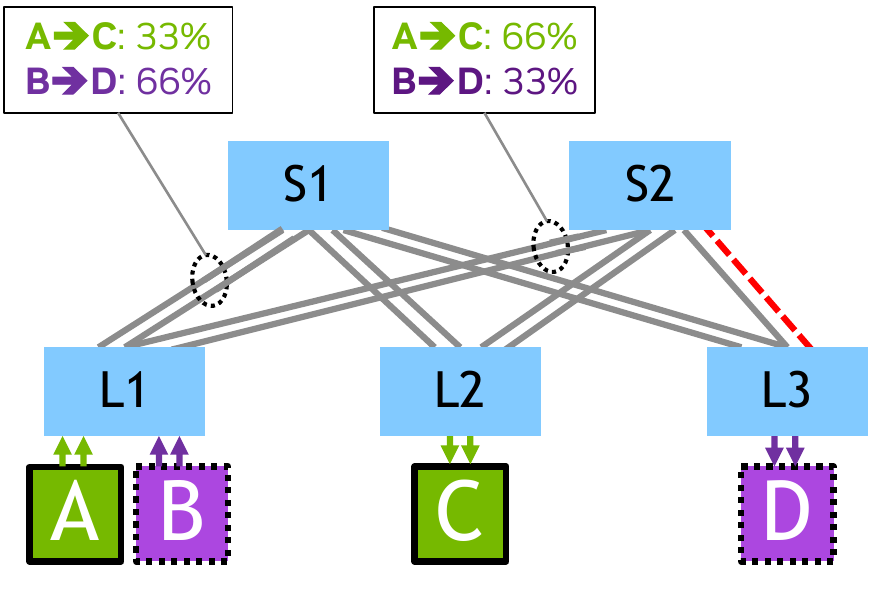}
    \caption{Asymmetrical load balancing. Weighted Adaptive Routing example in which remote destination D has reduced capacity. When traffic to D is active, the Adaptive Routing weights steer it towards S1, causing traffic to destination C to shift towards S2.}
    \label{fig:Weighted_AR}
\end{figure}

Spectrum switch adaptive routing reacts to local fabric link failures in O(100 ns) leveraging fully hardware-accelerated mechanisms. In contrast, when symmetry is broken due to non-local \emph{permanent} fabric failure, the response is handled via \emph{Weighted Adaptive Routing (AR)}, which accounts for the effective bandwidth capacity of remote path to destinations. The weights are computed by a BGP-based control plane \cite{ietf-bess-ebgp-dmz-10}, which tracks topology changes and distributes weighted routing state. Once installed, the hardware AR engine combines local queue depth with remote weights, enabling traffic to shift away from degraded regions at packet-balancing timescales as illustrated in \cref{fig:Weighted_AR}.

This separation allows fast, in-hardware reaction for local events, while relying on slower control-plane updates for global asymmetry.

\section{Lessons learned in large deployments}
%
Over two years of \projname production deployment, we have found \emph{end-to-end observability} across compute, collectives, NIC, and switch to be essential. Our observations are confirmed by correlating direct step-time degradation with the identified respective root cause. We share representative examples below.


\subsection{Debug experience of Adaptive Routing}

Hardware-based AR provides a unique operational advantage: it produces a structurally symmetric traffic pattern in the fabric. Because AR distributes traffic evenly across links, any deviation from a perfect balance serves as a sign of a problem.

\begin{figure}[htbp]
  \centering
  \includegraphics[width=\linewidth]{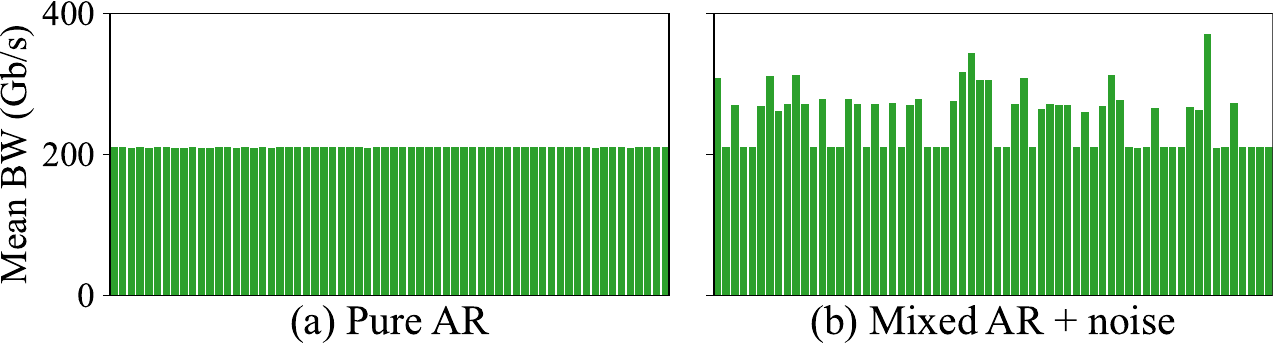}
  \caption{Leaf switch per-port uplink BW: (a) AR expected uniform pattern; (b) non-AR traffic interference.}
  \label{fig:ar_vs_mixed_load_balance}
\end{figure}

For example,  consider the per-port BW of uplinks in a particular leaf switch. \Cref{fig:ar_vs_mixed_load_balance}(a) shows a perfectly uniform traffic. (b) misconfigured NIC, switch, or collective lib, is immediately apparent by breaking the symmetry, as shown in \Cref{fig:ar_vs_mixed_load_balance}(b).

Uniformity anomalies detected across multiple so-called \emph{symmetry groups} such as leaf uplinks, rails and planes. Any outliers are likely indicators of hardware faults or software bugs.


\subsection{Catching stragglers via network indicators}
\label{ssec:histograms_stragglers}
In AI clusters, detecting anomalous low-performing nodes (stragglers) is crucial to maintaining high cluster utilization.
We have found that a large category of issues, whether networking-related or not, is detectable using networking tools only. This coarse-grained approach works because identifying stragglers is more time-critical than diagnosing their precise root causes.

\begin{figure}[!t]
    \centering
    \begin{subfigure}[b]{0.49\linewidth}
        \centering
        \raisebox{1.1ex}{%
            \includegraphics[width=\linewidth]{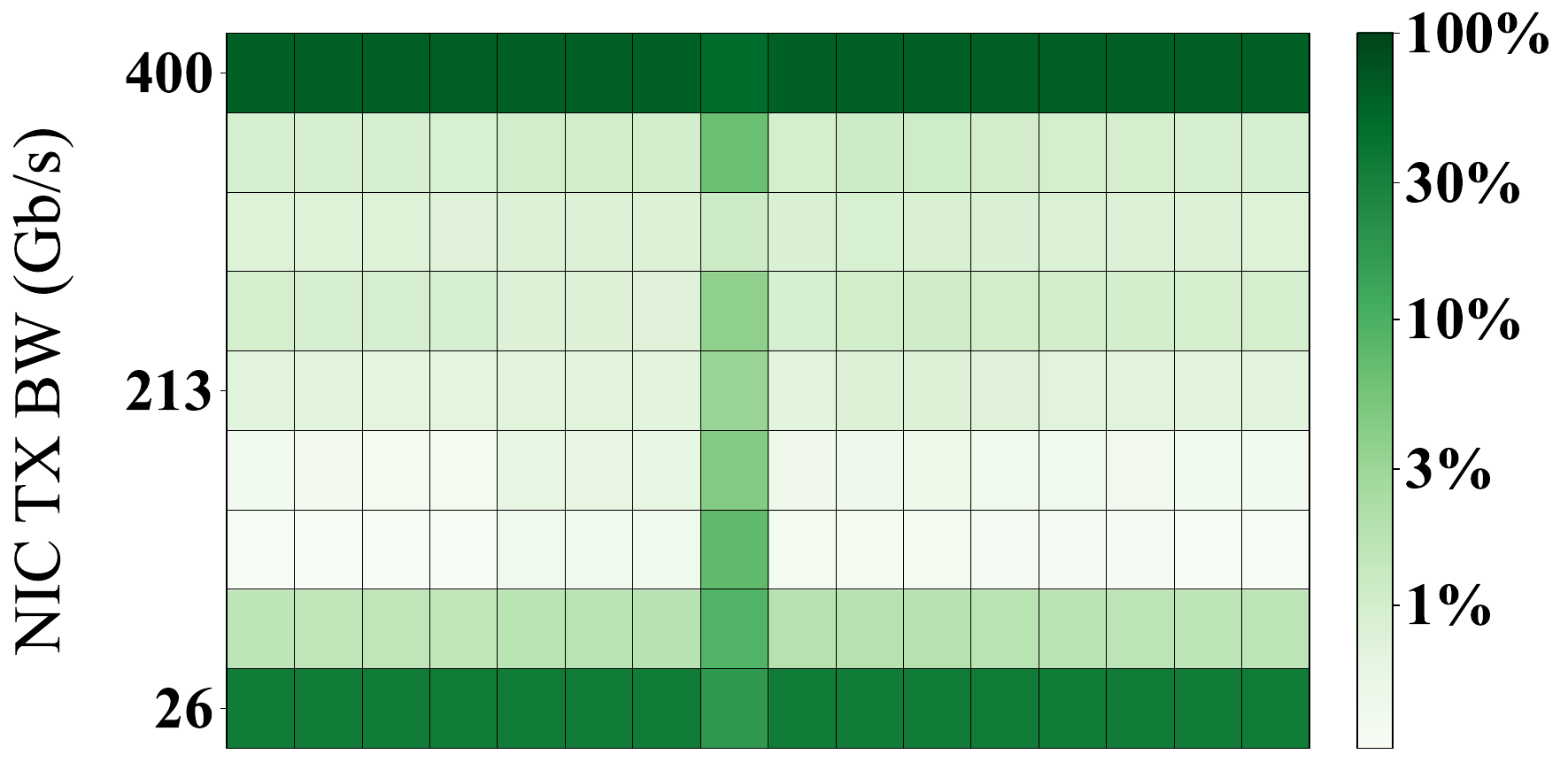}%
        }
        \caption{}
        \label{fig:straggler_bw_histogram}
    \end{subfigure}\hspace{0.01\linewidth}
    \begin{subfigure}[b]{0.49\linewidth}
        \centering
        \raisebox{0ex}{%
            \includegraphics[width=\linewidth]{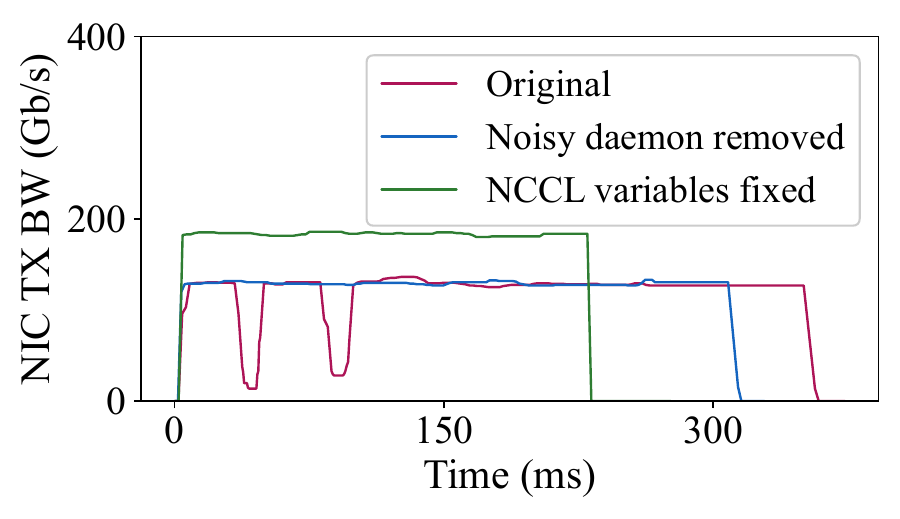}%
        }
        \caption{}
        \label{fig:noise_daemon}
    \end{subfigure}

    \caption{Debugging using high-frequency telemetry (HFT).
    \subref{fig:straggler_bw_histogram} Per-NIC BW histogram with straggler.
    \subref{fig:noise_daemon} HFT time-series debugging: daemon-induced BW drops and incorrect NCCL flags causing consistent BW under-utilization.}
    \label{fig:hft_debug_examples}
\end{figure}

Tightly-coupled collectives split large transfers into sequences of small, dependent messages - each only a few MB and transmitted in O(10\usec). A single straggler, fails to saturate the port BW and forces every other rank to stall at each message boundary. Consequently, while the straggler's BW fluctuates, all healthy participants exhibit a bi-modal BW distribution: either sending at line rate or idling. \projname's switch hardware exposes this intra-collective behavior through per-\usec histograms, as shown in \cref{fig:straggler_bw_histogram}.



\subsection{Network debugging with HFT}
\label{ssec:lessons_hft}
High-Frequency Telemetry (HFT) serves as a microscope for network traffic, exposing dynamics that standard polling intervals miss entirely. 
This visibility is essential for studying short-lived dynamics, such as transient queue build-up, bandwidth dips, and CC throttling, that coarse-grained counters cannot capture.
We generally consider streaming sampling intervals in the 100\usec to 10ms range as HFT. In addition to the previously presented histograms, \projname  support time-series HW accelerated HFT, from both NICs and switches, as demonstrated in this section.

For example, during the initial bringup, we discovered occasional drops in training benchmark performance (see \cref{fig:noise_daemon}, top). HFT plots have shown sharp transient BW drops during the collective, which led us to isolate a periodic process that was hogging system resources, and stop it. We also saw the NIC did not reach full line rate (marked as ``Noisy daemon removed'' in the graph); this led us to review the application configuration, which contained an error in passing NCCL flags.

HFT was also vital in tuning congestion-control to accommodate the needs of ever-evolving workloads and topologies. Using switch HFT we can view egress queues, BW, and PFCs, all on the same timeline and at high frequency. Such metrics are essential for tuning congestion control mechanisms.

\section{Evaluation}
\label{sec:evaluation}
We evaluate \projname on a variety of benchmarks and scenarios. The main evaluation results are summarized in \Cref{tab:eval_summary}.



\begin{table*}[t]
\centering
\footnotesize
\renewcommand{\arraystretch}{1.3}

\begin{tabular}{l l l l l}
\toprule
\textbf{Category} & \textbf{Test} & \textbf{Platform} & \textbf{Key Result}  & \textbf{Insight} \\
\midrule
\multirow{2}{*}{\shortstack[l]{Performance at\\High Utilization (\cref{ssec:performance_max_load})}}
  & RDMA bisection bandwidth & \multirow{2}{*}{HSP} & p01 BW is 98\% of line rate & \projname provides consistently \\
  & p99 latency under 75\% load  & & 8--9\,$\mu$s at 75\% load & high bandwidth and low latency \\
\midrule
\multirow{2}{*}{\shortstack[l]{Isolation (\cref{ssec:isolation})}}
  & NCCL \ATOA with noise & \multirow{2}{*}{BSP} & No degradation & \projname offers robust isolation \\
  & DeepSeek-V3 training with noise & & Stable 668\,ms/step & across concurrent workloads\\
\midrule
\multirow{2}{*}{\shortstack[l]{Static\\Resiliency (\cref{ssec:static_resiliency})}}
  & RDMA under 10\% link failures & \multirow{2}{*}{HSP} & $-11\%$ BW, $+7\%$ p99 LAT & \projname's performance scales  \\
  & NCCL Collectives, 10\% failures  & & 3-10\% from ideal & proportionally to hardware failures \\
\midrule
\multirow{2}{*}{\shortstack[l]{Dynamic\\Resiliency (\cref{ssec:dynamic_resiliency})}}
  & Single host link flap recovery & \multirow{2}{*}{BUMP} &  <3\,ms recovery & \projname achieves millisecond-scale \\
  & LLM training during link flap & & No impact (fabric), <5\% (host) & reaction time to failures\\
\midrule
\multirow{2}{*}{\shortstack[l]{Large-Scale\\Resiliency (\cref{ssec:large_scale_Resiliency})}}
  & Fabric link flap & \multirow{2}{*}{NSX} & No visible P99 CCT impact & \projname maintains, at scale, close to \\
  & Host link flap & & Fast convergence is crucial & optimal performance despite failures \\
\midrule
\multirow{2}{*}{\shortstack[l]{Multiplane\\Load Balancing (\cref{ssec:multiplane})}}
  & Asymmetry: one-to-many, \ATOA & \multirow{2}{*}{BUMP} & \projname: $<$3\% degradation & \projname preserves fault-free performance \\
  & State per-plane vs. per-path & & \projname ${\sim}$180\,GB/s at 4\,GB & under dynamic asymmetry \\
\bottomrule
\end{tabular}
\caption{\projname evaluation (HSP = Hopper\_SP, BSP = Blackwell\_SP, BUMP = Blackwell\_Ultra\_MP, NSX = NSX simulator).}
\label{tab:eval_summary}
\end{table*}

\subsection{Methodology}
\label{ssec:methodology}

\paragraph{Hardware Setup}
We evaluate the performance of \projname on three generations of clusters  (see \Cref{tab:cluster_configs}).

\begin{table}[t]
\centering
\small
\resizebox{\columnwidth}{!}{
\begin{tabular}{lccccc}
\toprule
\textbf{Cluster} & \textbf{GPUs} & \textbf{Nodes} & \textbf{Planes} & \textbf{Topology} & \textbf{NIC} \\
\midrule
Hopper\_SP & 1024 & 128 & 1 & 3LFT (rail-opt) & CX7 \\
Blackwell\_SP & 144 & 36 & 1 & 2LFT & CX7 \\
Blackwell\_Ultra\_MP & 1152 & 288 & 4 & 2LFT(rail-opt) & CX8 \\
\bottomrule
\end{tabular}
}
\caption{Cluster configurations used in evaluation.}
\label{tab:cluster_configs}
\end{table}

Topologies are scaled down by replacing many sparsely populated switches with fewer fully populated ones connected by parallel links; for instance, 100 spines at 10\% population become 10 fully populated spines with 10 parallel links, reducing hardware cost while preserving bisection bandwidth.
This consolidation mirrors real sub-max-scale production deployments.
It introduces additional challenges when evaluating link failures, which weighted-AR accounts for (\cref{sec:design:fault_tolerance}).

\paragraph{Simulation}
We use the NSX simulator~\cite{nsx} to evaluate network behavior at large scale.
NSX is an event-driven, GPU-accelerated network simulator that models the core technologies of \projname,
including multi-plane, adaptive routing, and congestion control.
Its results are continuously validated against smaller development clusters,
showing strong correlation in key performance metrics.


\paragraph{Workloads and scenarios}
We define our experiments as combinations of workloads (\textit{what} we test) and scenarios (\textit{how} we test).
Workloads cover: \textbf{RDMA bisection} microbenchmarks;
\textbf{NCCL collectives} (e.g., \AG, \AR, \ATOA); and \textbf{AI model} training.
We apply these across three scenarios:
\textbf{Baseline} (standalone, optimal performance),
\textbf{Isolation} (victim workload under concurrent background traffic),
and \textbf{Resiliency} (performance under emulated faults such as link failures or flaps).

%

Due to operational constraints, we do not evaluate all workloads and scenarios on all setups.
Table ~\ref{tab:eval_summary} summarizes the combinations presented in this section.
Clusters \textbf{Hopper\_SP} and \textbf{Blackwell\_SP} run an identical networking stack, which is also identical to the single-plane solution of \textbf{Blackwell\_Ultra\_MP} that adds multiplane load balancing.
When running workloads that do not use \emph{all} nodes,
we trim down the fabric links  to get a 1:1 symmetric non-blocking topology in order to not evaluate an under-subscribed network.

\paragraph{Metrics} We quantify performance using metrics tailored to each workload, focusing on measuring tail performance as discussed in \S\ref{sec:motivation}, unless stated otherwise. 
For NCCL tests, we evaluate the \emph{bus bandwidth} across varying message sizes. Bus bandwidth ~\cite{nccl_tests_perf} is a collective-agnostic metric that normalizes inter-GPU communication speed, allowing us to compare achieved throughput directly against the theoretical hardware capacity, independent of the number of participating GPUs.
The message size reflects the total data size that the collective operates on. 
Finally, for end-to-end AI workloads, we report the measured training step time.


\paragraph{Reference Solutions} In some experiments  we compare the performance of \textbf{\projname} against a standard \textbf{Ethernet (ETH)} baseline representing traditional RoCEv2 deployments.
The ETH configuration uses DCQCN congestion control and conventional ECMP.


\subsection{Performance under high utilization}
\label{ssec:performance_max_load}

We evaluate the maximum sustained bandwidth  and compare it to ETH.
We stress-test a 64-node subset of the Hopper\_SP cluster with a RDMA bisection benchmark.
We measure the achieved bandwidth across GPU  pairs across all participating nodes in a worst-case allocation pattern that forces all traffic to traverse a spine. 
%
%
%

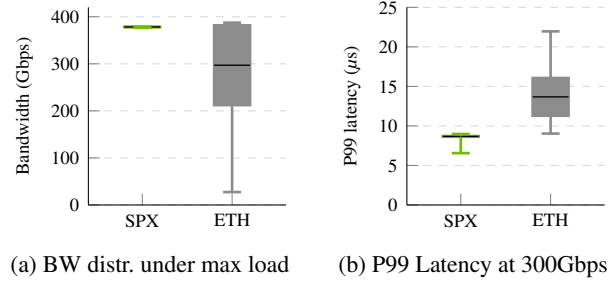
\begin{figure}[t]
    \centering
    \begin{subfigure}{0.236\textwidth}
        \centering
        \begin{tikzpicture}
\begin{axis}[
  width=\linewidth,  
  height=\linewidth, 
  ymin=0,
  ymax=420, 
  xmin=0.4,
  xmax=2.6,
  xtick={1,2},
  xticklabels={\projname,ETH},
  ytick={0,100,200,300,400},
  ylabel={Bandwidth (Gbps)},
  axis x line*=bottom,
  axis y line*=left,
  ymajorgrids=true,
  grid style={dashed, gray!30},
  tick label style={font=\scriptsize},      
  label style={font=\scriptsize},     
  legend style={
    font=\tiny,                       
    at={(0.03,0.05)},
    anchor=south west,
    draw=gray!50,
    fill=white,
    fill opacity=0.9,
    text opacity=1,
  },
]

\filldraw [fill=nvidia_green, draw=nvidia_green, line width=1pt] (0.8, 378.017) rectangle (1.2, 378.53);
\draw [black, line width=0.5pt] (0.8, 378.2735) -- (1.2, 378.2735);
\draw [nvidia_green, line width=1pt] (1, 378.017) -- (1, 376.59);
\draw [nvidia_green, line width=1pt] (0.9, 376.59) -- (1.1, 376.59);
\draw [nvidia_green, line width=1pt] (1, 378.53) -- (1, 378.56);
\draw [nvidia_green, line width=1pt] (0.9, 378.56) -- (1.1, 378.56);

\filldraw [fill=medium_gray, draw=medium_gray, line width=1pt] (1.8, 212.233) rectangle (2.2, 381.815);
\draw [black, line width=0.5pt] (1.8, 297.024) -- (2.2, 297.024);
\draw [medium_gray, line width=1pt] (2, 212.233) -- (2, 27.39);
\draw [medium_gray, line width=1pt] (1.9, 27.39) -- (2.1, 27.39);
\draw [medium_gray, line width=1pt] (2, 381.815) -- (2, 387.46);
\draw [medium_gray, line width=1pt] (1.9, 387.46) -- (2.1, 387.46);

\end{axis}
\end{tikzpicture}
        \subcaption{BW distr. under max load}
        \label{subfig:sol_spx_ots_bw}
    \end{subfigure}\hfill
    \begin{subfigure}{0.236\textwidth}
        \centering
        \begin{tikzpicture}
\begin{axis}[
  width=\linewidth,  
  height=\linewidth, 
  ymin=0,
  ymax=25,
  xmin=0.4,
  xmax=2.6,
  xtick={1,2},
  xticklabels={\projname,ETH},
  ytick={0,5,10,15,20,25},
  ylabel={P99 latency ($\mu$s)},
  axis x line*=bottom,
  axis y line*=left,
  ymajorgrids=true,
  grid style={dashed, gray!30},
  tick label style={font=\scriptsize},      
  label style={font=\scriptsize},     
  legend style={
    font=\tiny,                       
    at={(0.03,0.97)},
    anchor=north west,
    draw=gray!50,
    fill=white,
    fill opacity=0.9,
    text opacity=1,
  },
]

\filldraw [fill=nvidia_green, draw=nvidia_green, line width=1pt] (0.8, 8.630) rectangle (1.2, 8.720);
\draw [black, line width=0.5pt] (0.8, 8.675) -- (1.2, 8.675);
\draw [nvidia_green, line width=1pt] (1, 8.630) -- (1, 6.560);
\draw [nvidia_green, line width=1pt] (0.9, 6.560) -- (1.1, 6.560);
\draw [nvidia_green, line width=1pt] (1, 8.720) -- (1, 8.980);
\draw [nvidia_green, line width=1pt] (0.9, 8.980) -- (1.1, 8.980);

\filldraw [fill=medium_gray, draw=medium_gray, line width=1pt] (1.8, 11.290) rectangle (2.2, 16.062);
\draw [black, line width=0.5pt] (1.8, 13.676) -- (2.2, 13.676);
\draw [medium_gray, line width=1pt] (2, 11.290) -- (2, 9.030);
\draw [medium_gray, line width=1pt] (1.9, 9.030) -- (2.1, 9.030);
\draw [medium_gray, line width=1pt] (2, 16.062) -- (2, 21.960);
\draw [medium_gray, line width=1pt] (1.9, 21.960) -- (2.1, 21.960);

\end{axis}
\end{tikzpicture}
        \subcaption{P99 Latency at 300Gbps}
        \label{subfig:sol_spx_ots_lat}
    \end{subfigure}
    \caption{Performance under load: \projname vs. Ethernet (ETH).}
    \label{fig:sol_spx_ots}
\end{figure}

\Cref{subfig:sol_spx_ots_bw} shows ETH's bandwidth variability across pairs, with the median at 300~Gbps and some pairs collapsing to 25~Gbps, due to ECMP's static flow hashing.
Conversely, \projname achieves a tight distribution. 
With a 1st percentile bandwidth of 377.23~Gbps (98\% of the theoretical line rate), \projname ensures predictable, sustained throughput across all GPU pairs.

\Cref{subfig:sol_spx_ots_lat} compares P99 latency across pairs for both \projname and ETH at 300~Gbps (rate-limiting to match the median bandwidth sustained by ETH).  
We run \texttt{ib\_send\_lat} concurrently with \texttt{ib\_write\_bw} background traffic.
ETH achieves the median P99 latency among pairs of $\sim$13~\usec, with a very broad spread reaching 22~\usec.
These delays translate into severe bottlenecks for AI workloads (\S\ref{ssec:isolation}). In contrast, \projname maintains a low, tightly clustered p99 latency with a median of 8--9~\usec, showing  stable performance under load.


\subsection{Performance isolation}
\label{ssec:isolation}


\begin{figure}[t]
    \centering 
    \begin{subfigure}{0.495\linewidth} 
        \centering
        \begin{tikzpicture}
  \begin{axis}[
  nccl axis,
  xmode=log,
  width=\linewidth,      
  height=\linewidth, 
  ymax=60,
  ymin=0,                
  xmax=4.294967296e9,    
    xtick={131072, 33554432, 4.294967296e9},
    xticklabels={128 KB, 32 MB, 4 GB},
  tick label style={font=\scriptsize},
  label style={font=\footnotesize},
  legend style={
    font=\scriptsize,
    at={(0.02,0.98)},    
    anchor=north west
  },
]

  \addplot[
    color=medium_gray,
    line width=1.75pt,
    mark=square*,
    mark size=1.5pt,
    mark options={solid, fill=white, draw=medium_gray, line width=1.2pt},
  ] coordinates {
    (131072,0.39) (262144,1.22) (524288,3.22) (1048576,6.01) (2097152,10.97) (4194304,17.88) (8388608,24.77) (16777216,31.68) (33554432,34.09) (67108864,39.5) (134217728,40.43) (268435456,41.49) (536870912,42.17) (1073741824,42.39) (2147483648,43.08) (4.29497e9,43.47)
  };
  \addlegendentry{ETH}

  \addplot[
    color=nvidia_green,
    line width=1.75pt,
    mark=*,
    mark size=1.5pt,
    mark options={solid, fill=nvidia_green},
  ] coordinates {
    (131072,0.33) (262144,1.28) (524288,3.18) (1048576,6.24) (2097152,10.67) (4194304,17.54) (8388608,24.3) (16777216,33.03) (33554432,35.15) (67108864,40.32) (134217728,45.73) (268435456,47.38) (536870912,48.23) (1073741824,48.75) (2147483648,49.16) (4.29497e9,49.38)
  };
  \addlegendentry{\projname}

  \end{axis}
\end{tikzpicture}
        \label{subfig:nccl_isolation_baseline}
    \end{subfigure}        
    \begin{subfigure}{0.495\linewidth}
        \centering
        \begin{tikzpicture}
  \begin{axis}[
  nccl axis,
  xmode=log,
  width=\linewidth,      
  height=\linewidth, 
  ymax=60,
  ymin=0,                
  xmax=4.294967296e9,    
xtick={131072, 33554432, 4.294967296e9},
xticklabels={128 KB, 32 MB, 4 GB},
  tick label style={font=\scriptsize},
  label style={font=\footnotesize},
  legend style={
    font=\scriptsize,
    at={(0.02,0.98)},    
    anchor=north west
  },
]

  \addplot[
    color=medium_gray,
    line width=1.75pt,
    mark=square*,
    mark size=1.5pt,
    mark options={solid, fill=white, draw=medium_gray, line width=1.2pt},
  ] coordinates {
    (131072,0.27) (262144,0.94) (524288,2.16) (1048576,2.92) (2097152,5.25) (4194304,4.43) (8388608,10.37) (16777216,8.6) (33554432,8.56) (67108864,9.0) (134217728,9.27) (268435456,9.65) (536870912,10.12) (1073741824,10.24) (2147483648,10.55) (4.29497e9,10.93)
  };
  \addlegendentry{ETH}

  \addplot[
    color=nvidia_green,
    line width=1.75pt,
    mark=*,
    mark size=1.5pt,
    mark options={solid, fill=nvidia_green},
  ] coordinates {
    (131072,0.37) (262144,1.34) (524288,3.03) (1048576,5.25) (2097152,10.75) (4194304,16.61) (8388608,22.3) (16777216,31.61) (33554432,34.33) (67108864,41.39) (134217728,45.79) (268435456,47.45) (536870912,48.5) (1073741824,48.92) (2147483648,49.2) (4.29497e9,49.39)
  };
  \addlegendentry{\projname}

  \end{axis}
\end{tikzpicture}
        \label{subfig:nccl_isolation_noise}
    \end{subfigure}    
    \caption{Single \ATOA (left); two concurrent \ATOA (right).}
    \label{fig:nccl_isolation}
\end{figure}
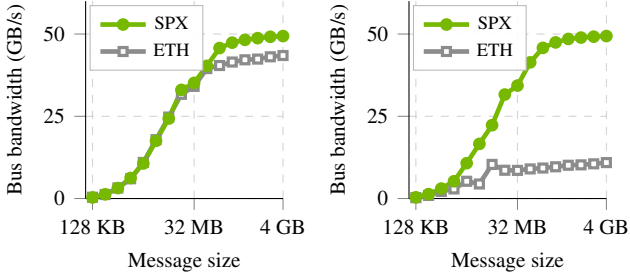

\Cref{fig:nccl_isolation} (left) shows that for isolated \ATOA collectives, ETH performs on par with \projname for small message sizes, but suffers at higher loads, peaking at 43~GB/s. In contrast, \projname achieves 49.3~GB/s ($1.13\times$),  nearly 99.5\% of the theoretical 49.5~GB/s hardware capacity. This aligns with  \S\ref{ssec:performance_max_load} as ETH cannot sustain maximum bandwidth under heavy load.
\Cref{fig:nccl_isolation} (right) shows  the impact of contention.
Here we execute two \ATOA collectives in parallel, allocating 16 nodes to the "victim" workload and 48 nodes to the background "noise" collective. 
 With ETH, the victim's bandwidth stagnates below 10.93~GB/s, about 80\% performance drop. In contrast, \projname achieves near-perfect isolation. 

\begin{figure}
\begin{tikzpicture}
  \begin{axis}[
    nccl axis,
    xmode=linear,
    width=0.43\textwidth,
    height=0.22\textwidth,
    ymin=0,
    ymax=1800,
    xmin=0,
    xmax=85,
    xlabel={Iteration},
    ylabel={Step time (ms)},
    xtick={0,20,40,60,80},
    xticklabels={0,20,40,60,80},
    ytick={0,300,600,900,1200,1500},
    legend style={
      at={(0.02,0.96)},
      anchor=north west
    },
    axis lines=left,
    ymajorgrids=true,
    xmajorgrids=true,
    grid style={dashed, gray!30},
    clip=false,
  ]
  
  \draw[dashed, black, line width=0.8pt] (axis cs:42.5,0) -- (axis cs:42.5,1800);
  
  \node[anchor=north, font=\small, text=black] at (axis cs:21,400) {Baseline};
  \node[anchor=north, font=\small, text=black] at (axis cs:63.5,400) {Isolation Test};
  
  \node[anchor=west, align=left, font=\small, text=nvidia_green] at (axis cs:45,1500) {\projname $\sim$2x faster \\ step time};
  
  \addplot[
    color=medium_gray,
    line width=1.5pt,
    solid,
    mark=none,
  ] coordinates {
    (1,666.6) (2,787.1) (3,661.2) (4,651.8) (5,703.2) (6,647.6) (7,644.9) (8,652.0) (9,645.4) (10,647.5) (11,647.3) (12,664.5) (13,648.3) (14,650.4) (15,669.8) (16,659.0) (17,658.9) (18,648.1) (19,650.6) (20,650.9) (21,642.4) (22,657.0) (23,660.5) (24,661.5) (25,646.5) (26,657.0) (27,654.0) (28,662.0) (29,655.5) (30,649.1) (31,647.3) (32,645.0) (33,642.4) (34,650.1) (35,645.1) (36,645.9) (37,644.8) (38,650.7) (39,655.0) (40,648.4) (41,651.2) (42,648.0) (43,953.7) (44,1106.3) (45,1097.7) (46,1105.2) (47,1087.9) (48,1137.8) (49,1098.5) (50,1070.2) (51,1100.4) (52,1166.1) (53,1118.4) (54,1103.2) (55,1107.1) (56,1087.3) (57,1075.2) (58,1087.2) (59,1077.5) (60,1071.6) (61,1082.4) (62,1095.7) (63,1068.3) (64,1085.1) (65,1113.4) (66,1117.5) (67,1226.7) (68,1135.7) (69,1132.9) (70,1091.1) (71,1085.6) (72,1084.6) (73,1104.5) (74,1098.6) (75,1089.2) (76,1138.7) (77,1116.6) (78,1086.1) (79,1056.3) (80,1042.0) (81,1106.8) (82,1102.4) (83,1082.4) (84,1096.1)
  };
  \addlegendentry{ETH}
  
  \addplot[
    color=nvidia_green,
    line width=1.5pt,
    solid,
    mark=none,
  ] coordinates {
    (1,640.0) (2,637.2) (3,633.3) (4,633.1) (5,632.2) (6,629.2) (7,630.9) (8,628.2) (9,628.8) (10,626.7) (11,631.3) (12,631.0) (13,632.2) (14,631.4) (15,632.8) (16,634.6) (17,631.5) (18,630.2) (19,629.4) (20,630.7) (21,629.7) (22,631.6) (23,631.2) (24,630.5) (25,631.5) (26,629.3) (27,630.1) (28,630.9) (29,631.0) (30,629.8) (31,630.4) (32,630.2) (33,631.0) (34,632.9) (35,633.0) (36,631.1) (37,632.5) (38,632.1) (39,632.3) (40,633.1) (41,634.0) (42,633.2) (43,640.6) (44,635.9) (45,634.5) (46,633.7) (47,634.3) (48,632.6) (49,633.3) (50,631.9) (51,633.8) (52,631.3) (53,632.4) (54,631.6) (55,630.9) (56,630.7) (57,631.4) (58,631.6) (59,631.2) (60,631.0) (61,631.0) (62,635.5) (63,629.5) (64,629.7) (65,630.5) (66,629.2) (67,632.9) (68,630.9) (69,631.0) (70,630.4) (71,632.1) (72,629.9) (73,631.5) (74,632.0) (75,629.7) (76,632.6) (77,635.3) (78,631.6) (79,631.0) (80,630.3) (81,630.8) (82,633.4) (83,634.8) (84,634.1)
  };
  \addlegendentry{\projname}
  
  \end{axis}
  \end{tikzpicture}
  
\caption{DeepSeek-V3 Isolation 16N NVL8 proxy model.}
\label{fig:dsv3_comparison}
\end{figure}
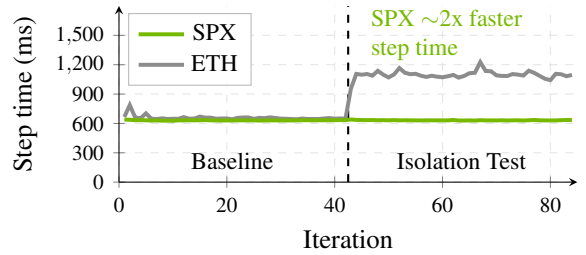

\paragraph{Impact of background load on training step times} 
We run a DeepSeek-V3 LLM model training session on a 16-node subset of the Blackwell\_SP cluster. The LLM nodes are distributed randomly uniform across the cluster. As shown in \cref{fig:dsv3_comparison}, in the standalone run \projname achieves an average training step time of 667~ms, whereas ETH averages 735~ms. Under background load of RDMA bisection noise on an additional 16 nodes,
ETH suffers severe degradation, with step times increasing by $1.6\times$ to 1.18~s. Conversely, \projname  maintains the same step time of 668~ms, demonstrating robust performance isolation.

\subsection{Static resiliency}
\label{ssec:static_resiliency}

We aim to show that the network performance degrades proportionally to the loss of bisection capacity. We configure a 60-node subset of the Hopper\_SP cluster in a trimmed topology. We compare a failure-free, pristine baseline against a degraded network with leaf-to-spine links that are permanently disabled matching the max-flow capacity seen between different leaf pairs at early stages of a cluster's life (\S~\ref{ssec:time-to-ai}).

We execute two types of workloads. First, to analyze the impact of failures under maximum utilization, we reproduce the RDMA bisection test from \S\ref{ssec:performance_max_load}. We run \texttt{ib\_write\_bw} across GPU pairs in the same rail to generate bisection traffic, while concurrently measuring the p99 tail latency across pairs using \texttt{ib\_send\_lat}. Second, we run NCCL \ATOA collective benchmark from \cref{ssec:time-to-ai}.

\paragraph{RDMA}
Our results show that under the degraded fabric with 10\% less links, the p01 bandwidth gracefully degrades by 11\% from 377.80 Gbps to 335.31 Gbps, closely tracking the actual physical network capacity. Similarly, the p99 latency shows an increase from 14.97 $\mu$s to 15.96 $\mu$s.


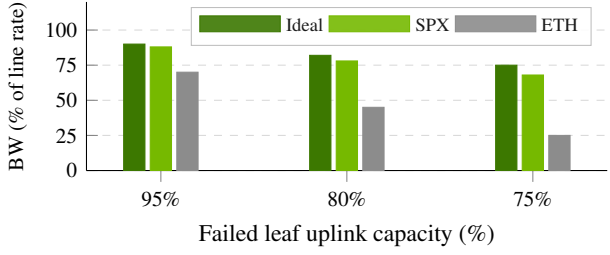
\begin{figure}
  \centering
  \begin{tikzpicture}
  \begin{axis}[
    ybar,
    width=\linewidth,
    height=0.45\linewidth,
    bar width=8pt,
    ymin=0, ymax=120,
    ytick={0, 25, 50, 75, 100},
    ylabel={BW (\% of line rate)},
    symbolic x coords={95pct, 80pct, 75pct},
    xtick={95pct, 80pct, 75pct},
    xticklabels={95\%, 80\%, 75\%},
    xlabel={Failed leaf uplink capacity (\%)},
    axis x line*=bottom,
    axis y line*=left,
    ymajorgrids=true,
    grid style={dashed, gray!30},
    enlarge x limits=0.2,
    tick label style={font=\footnotesize},
    xlabel style={font=\small},
    ylabel style={font=\footnotesize, align=center},
    legend image code/.code={\fill[#1] (0,-0.07cm) rectangle (0.08cm,0.14cm);},
    legend style={
      font=\scriptsize,
      at={(0.97,0.97)},
      anchor=north east,
      fill=white,
      fill opacity=0.9,
      draw=gray!50,
      row sep=2pt,
    },
    legend columns=3,
  ]

  \addplot[fill=dark_green, draw=dark_green] coordinates {
    (95pct,90) (80pct,82) (75pct,75)
  };
  \addlegendentry{Ideal}

  \addplot[fill=nvidia_green, draw=nvidia_green] coordinates {
    (95pct,88) (80pct,78) (75pct,68)
  };
  \addlegendentry{\projname}

  \addplot[fill=medium_gray, draw=medium_gray] coordinates {
    (95pct,70) (80pct,45) (75pct,25)
  };
  \addlegendentry{ETH}

  \end{axis}
\end{tikzpicture}
  \caption{\ATOA performance under blocking fabric.}
  \label{fig:perf_under_failures_with_spx}
\end{figure}

\paragraph{NCCL collectives} 
\Cref{fig:perf_under_failures_with_spx} depicts link failure percentages and the performance impact in an \ATOA collective, as described in \Cref{ssec:time-to-ai}.
While traditional Ethernet solutions degrade in a non-proportional way to the bandwidth loss, \projname maintains 3-10\% of the ideal solution bandwidth.



\subsection{Dynamic resiliency}
\label{ssec:dynamic_resiliency}

We compare the performance of \projname's hardware-accelerated Plane Load Balancer (PLB) to a software-based NCCL reference solution (SW LB) and a single-plane configuration.

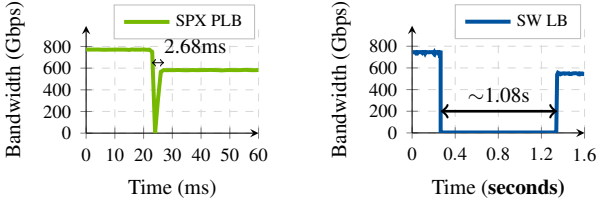
\begin{figure}[t]
    \centering 
    \begin{subfigure}{.48\linewidth} 
        \centering
        \begin{tikzpicture}
    \begin{axis}[
      xmode=linear,
      width=0.95\linewidth, 
      height=3.0cm,    
      ymin=0,
      ymax=990,
    xmin=0,
    xmax=60,
    xlabel={Time (ms)},
    ylabel={Bandwidth (Gbps)},
    xtick={0, 10, 20, 30, 40, 50, 60},
      /pgf/number format/1000 sep={}, 
      legend style={
        font=\fontsize{9}{10}\selectfont,
        draw=gray!50,
        at={(0.98,1.2)},
        anchor=east
      },
      axis lines=left,
      ymajorgrids=true,
      xmajorgrids=true,
      grid style={dashed, gray!30},
      clip=true, 
      tick label style={font=\scriptsize},
      label style={font=\footnotesize},
    ]
    
    \addplot[
      color=nvidia_green,
      line width=1.5pt,
      solid,
      mark=none,
    ] table [x expr=\thisrow{timestamp_ms}-7360, y=bw_gbps, col sep=tab] {figures/reaction_time/hwplb.csv};
    \addlegendentry{\footnotesize\projname PLB}
  
    \draw[<->, black, line width=0.1pt] (axis cs:23,650) -- (axis cs:27,650);
    \node[anchor=south, align=left, font=\footnotesize, text=black] at (axis cs:38,650) {2.68ms};
    
    \end{axis}
  \end{tikzpicture}
        \label{subfig:reaction_time_hwplb}
    \end{subfigure}
    ~
    \begin{subfigure}{.48\linewidth} 
        \centering
        \begin{tikzpicture}
    \begin{axis}[
      xmode=linear,
      width=0.95\linewidth, 
      height=3.0cm,        
      ymin=0,
      ymax=990,
    xmin=0,
    xmax=1.6,
    xlabel={Time (\bf{seconds})},
    ylabel={Bandwidth (Gbps)},
    xtick={0, 0.4, 0.8, 1.2, 1.6},
      /pgf/number format/1000 sep={},
      legend style={
        font=\fontsize{9}{10}\selectfont,
        draw=gray!50,
        at={(0.98,1.2)},
        anchor=east
      },
      axis lines=left,
      ymajorgrids=true,
      xmajorgrids=true,
      grid style={dashed, gray!30},
      clip=true, 
      tick label style={font=\scriptsize},
      label style={font=\footnotesize},
    ]
    
    \addplot[
      color=dark_blue,
      line width=1.5pt,
      solid,
      mark=none,
    ] table [x expr=\thisrow{time_s}-0.4, y=bw_gbps, col sep=tab] {figures/reaction_time/swlb.csv}; 
    \addlegendentry{\footnotesize SW LB}
  
    \draw[<->, black, line width=1.0pt] (axis cs:0.265,200) -- (axis cs:1.346,200);
    \node[anchor=south, align=center, font=\footnotesize, text=black] at (axis cs:0.805,200) {$\sim$1.08s};
    
    \end{axis}
  \end{tikzpicture}
        \label{subfig:reaction_time_swlb}
    \end{subfigure}%
    \caption{Endpoint single flap recovery in Blackwell\_Ultra\_MP with \projname hardware multi-plane (PLB) vs. software load balancer (SW LB).}
    \label{fig:reaction_time}
\end{figure}

\paragraph{Recovery time for a single host link flap} We execute \texttt{ib\_write\_bw} between node pairs and inject a link failure in a single host link, failing one plane. 
In the single-plane case, the RDMA connection crashes and communication halts entirely. \Cref{fig:reaction_time} (left) shows the plane failover under \projname. The flow initially sustains line-rate bandwidth. Upon link failure, bandwidth drops to zero as packets on the failed plane are lost, triggering retransmissions. However, in under 3~ms, PLB  redistributes traffic across the three other planes and reaches 75\% of the original line rate. Since the software load balancer operates above the NCCL layer, we test its performance using a \SR benchmark across node pairs. \Cref{fig:reaction_time} (right) shows that SW LB requires 1.08~\emph{seconds} to recover, about $400\times$ slower than hardware PLB.

\begin{figure}[t]
    \centering 
    \begin{tikzpicture}
  \begin{axis}[
    width=\linewidth,
    height=0.4\linewidth, 
    ymin=2900,
    ymax=3200,
    xmin=0,
    xmax=154, 
    xlabel={Iteration},
    ylabel={Step Time (s)}, 
    axis lines=left,
    ymajorgrids=true,
    xmajorgrids=true,
    grid style={dashed, gray!30},
    clip=false,
    tick label style={font=\scriptsize}, 
    label style={font=\scriptsize},
    xtick distance=20, 
    ytick distance=50,
    scaled y ticks=false,
    yticklabel={%
      \pgfmathparse{\tick/1000}%
      \pgfmathprintnumber[fixed, precision=2, zerofill]{\pgfmathresult}%
    },
    legend columns=2, 
    legend style={
      at={(0.6,1.20)}, 
      draw=gray!50,
      anchor=north,
      font=\scriptsize, 
      cells={anchor=west}, 
      /tikz/every even column/.append style={column sep=0.2cm} 
    },
  ]
  
  \draw[dashed, black, opacity=0.5] (axis cs:0,2950) -- (axis cs:154,2950);
  \draw[dashed, black, opacity=0.5] (axis cs:0,3050) -- (axis cs:154,3050);

  \fill[gray!30] (axis cs:6,2900) rectangle (axis cs:18,3200);
  \fill[gray!30] (axis cs:25,2900) rectangle (axis cs:38,3200);

  \fill[green!50!black!20] (axis cs:71,2900) rectangle (axis cs:80,3200);
  \fill[green!50!black!20] (axis cs:110,2900) rectangle (axis cs:119,3200);
  \fill[green!50!black!20] (axis cs:134,2900) rectangle (axis cs:145,3200);

  \addplot[
    color=nvidia_green!60, 
    line width=0.5pt,
    solid,
    mark=*,
    mark size=0.5pt,
  ] coordinates {
    (0,2960.6) (1,2965.9) (2,2976.0) (3,3062.8) (4,2965.1) (5,3020.8) (6,3003.0) (7,3064.0) (8,3084.3) (9,3066.9) (10,3093.1) (11,3078.3) (12,3115.2) (13,3066.7) (14,3066.4) (15,3060.2) (16,3174.2) (17,3123.0) (18,3017.8) (19,2994.0) (20,2986.2) (21,2971.9) (22,2968.4) (23,2967.9) (24,2970.4) (25,3023.2) (26,3059.7) (27,3081.2) (28,3071.2) (29,3068.8) (30,3065.6) (31,3189.9) (32,3065.2) (33,3066.8) (34,3069.9) (35,3067.9) (36,3076.7) (37,3073.3) (38,3070.5) (39,3064.5) (40,3052.7) (41,2960.8) (42,2958.9) (43,2968.2) (44,2956.8) (45,2964.5) (46,2961.6) (47,2964.2) (48,2966.9) (49,2954.6) (50,2958.5) (51,2967.1) (52,2958.9) (53,2953.7) (54,2960.8) (55,2955.5) (56,2960.5) (57,2958.6) (58,2957.7) (59,2961.2) (60,2962.1) (61,2962.6) (62,2959.4) (63,2957.8) (64,2961.9) (65,2954.3) (66,2964.0) (67,2967.4) (68,2958.3) (69,3056.5) (70,3027.1) (71,2961.9) (72,2963.9) (73,2965.1) (74,2961.2) (75,3038.7) (76,2962.7) (77,3042.4) (78,2961.7) (79,2961.1) (80,2953.6) (81,2959.6) (82,2960.9) (83,2959.4) (84,3006.0) (85,2958.4) (86,2984.3) (87,2960.1) (88,2954.9) (89,2960.5) (90,2956.3) (91,2959.9) (92,2956.9) (93,3057.6) (94,2955.3) (95,2953.6) (96,2969.5) (97,2960.9) (98,2960.2) (99,2958.5) (100,2951.5) (101,2951.4) (102,2956.2) (103,2954.7) (104,2959.4) (105,2956.9) (106,2949.1) (107,2951.3) (108,2953.3) (109,2950.8) (110,2961.1) (111,2957.5) (112,2979.2) (113,3059.6) (114,2956.8) (115,2997.2) (116,2955.6) (117,2954.3) (118,2948.4) (119,2967.3) (120,2970.1) (121,2954.5) (122,2956.3) (123,2956.1) (124,2963.4) (125,2956.6) (126,2959.9) (127,2955.9) (128,2957.4) (129,2953.4) (130,2952.6) (131,3072.5) (132,2954.8) (133,2958.8) (134,2955.1) (135,2952.3) (136,2950.1) (137,2956.9) (138,2975.2) (139,3033.9) (140,2955.9) (141,2966.6) (142,2958.3) (143,2953.1) (144,2989.4) (145,2958.8) (146,2955.2) (147,2970.7) (148,2957.4) (149,2956.4) (150,3008.6) (151,2959.4) (152,2965.8) (153,2957.0)
  };
  \addlegendentry{Original Step Time}

  \addplot[
    color=nvidia_green,
    line width=1.5pt,
    solid,
    mark=none,
  ] coordinates {
    (0,2960.6) (1,2963.2) (2,2967.5) (3,2991.3) (4,2986.1) (5,2998.1) (6,3005.5) (7,3023.1) (8,3027.4) (9,3047.8) (10,3062.3) (11,3077.3) (12,3087.6) (13,3084.0) (14,3083.9) (15,3077.4) (16,3096.5) (17,3098.1) (18,3088.3) (19,3073.8) (20,3059.0) (21,3018.6) (22,2987.7) (23,2977.7) (24,2973.0) (25,2980.4) (26,2997.9) (27,3020.5) (28,3041.1) (29,3060.8) (30,3069.3) (31,3095.3) (32,3092.1) (33,3091.3) (34,3091.5) (35,3091.9) (36,3069.3) (37,3070.9) (38,3071.7) (39,3070.6) (40,3067.5) (41,3044.4) (42,3021.5) (43,3001.0) (44,2979.5) (45,2978.0) (46,2967.9) (47,2966.0) (48,2965.2) (49,2962.4) (50,2961.2) (51,2962.3) (52,2961.2) (53,2958.6) (54,2959.8) (55,2959.2) (56,2957.9) (57,2957.8) (58,2958.6) (59,2958.7) (60,2960.0) (61,2960.4) (62,2960.6) (63,2960.6) (64,2960.8) (65,2959.2) (66,2959.5) (67,2961.1) (68,2961.2) (69,2980.8) (70,2979.8) (71,2980.1) (72,2974.4) (73,2975.6) (74,2975.8) (75,2978.2) (76,2978.3) (77,2994.0) (78,2993.3) (79,2993.3) (80,2976.3) (81,2975.7) (82,2959.4) (83,2958.9) (84,2967.9) (85,2968.9) (86,2973.8) (87,2973.6) (88,2972.7) (89,2963.6) (90,2963.2) (91,2958.3) (92,2957.7) (93,2978.2) (94,2977.2) (95,2976.7) (96,2978.6) (97,2979.4) (98,2959.9) (99,2960.5) (100,2960.1) (101,2956.5) (102,2955.6) (103,2954.5) (104,2954.6) (105,2955.7) (106,2955.3) (107,2954.3) (108,2954.0) (109,2952.3) (110,2953.1) (111,2954.8) (112,2960.4) (113,2981.6) (114,2982.8) (115,2990.1) (116,2989.7) (117,2984.7) (118,2962.5) (119,2964.6) (120,2959.1) (121,2958.9) (122,2959.3) (123,2960.9) (124,2960.1) (125,2957.4) (126,2958.5) (127,2958.4) (128,2958.6) (129,2956.6) (130,2955.8) (131,2978.4) (132,2978.1) (133,2978.4) (134,2978.8) (135,2978.7) (136,2954.2) (137,2954.6) (138,2957.9) (139,2973.7) (140,2974.4) (141,2977.7) (142,2978.0) (143,2973.6) (144,2964.7) (145,2965.2) (146,2963.0) (147,2965.4) (148,2966.3) (149,2959.7) (150,2969.7) (151,2970.5) (152,2969.5) (153,2969.4)
  };
  \addlegendentry{Moving Average} 

  \addlegendimage{area legend, fill=gray!30, draw=gray!30}
  \addlegendentry{Host Flap}
  \addlegendimage{area legend, fill=green!50!black!20, draw=green!50!black!20}
  \addlegendentry{Fabric Flap}

  \end{axis}
\end{tikzpicture}
    \caption{Nemotron 3 Ultra training under dynamic failures.}
    \label{fig:dynamic_failures}
\end{figure}
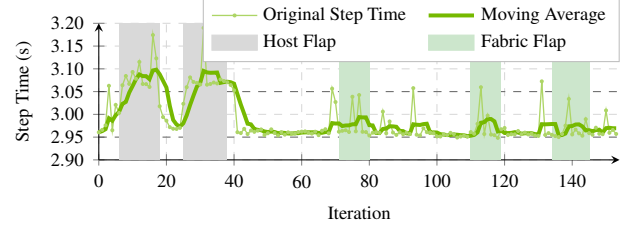

\paragraph{Link flaps impact on LLM training} 
We inject link flaps during a training job and measure the impact on step time. We train Nemotron 3 Ultra on a 64-node subset of the Blackwell\_Ultra\_MP cluster. \Cref{fig:dynamic_failures} shows the measured step times. We first inject two consecutive failures on a host-to-leaf link, affecting one of the four planes (gray shaded areas). \projname falls back to the three healthy planes in a single iteration. Once the link recovers, \projname instantly restores traffic to all four planes, returning to the 2.95s baseline. We later inject \emph{three independent flaps} on a leaf-to-spine uplink (green shaded regions). These have a negligible impact on step time stability, showing \projname's robust dynamic resiliency also at the fabric-tier.

\subsection{Large-scale resiliency}
\label{ssec:large_scale_Resiliency}

In this section we evaluate the expected \projname performance under failures in massive-scale clusters. As we do not have access to such clusters, we simulate them using NSX \cite{nsx}. 

\paragraph{Fabric flaps}
 We generate random link flaps using a fixed value of Mean Time Between Failures (MTBF) per link corresponding to 10 flaps/minute in a 64K single plane two level fat tree cluster. Upon flap the respective switch port is locally disabled for a \textit{flap duration}, 10 seconds, and immediately re-enabled.
 The control plane is not aware of the flap. The  MTBF values and flap duration are derived from conservative worst-case estimates from real large-scale clusters deployments.  
The  workload is 256 Ring-AllGather/ReduceScatter collectives, each 256 ranks. We aim to estimate P99 CCT among all these collectives. This value represents a good proxy for the expected performance in a large-scale training session over all cluster GPUs.

The fabric experiences performance degradation as a function of the number of concurrent failed links due to the flap, which has a Poisson distribution defined by the flap frequency and duration. Consequently, we run the simulations of this workload while varying the number of concurrent link failures from 1 to 10, record the respective P99 CCT for each, and obtain the expected P99 CCT of each collective as a weighted sum of these values according to their probabilities. 
 
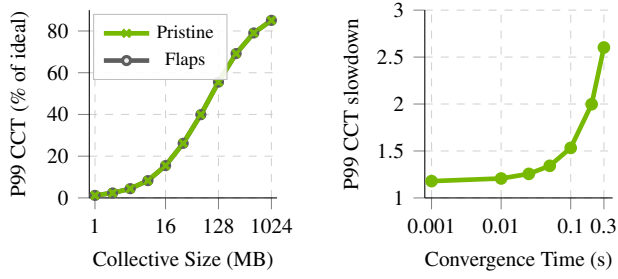
\begin{figure}
    \begin{subfigure}[t]{0.48\linewidth} 
        \begin{tikzpicture}[baseline=(current axis.north)]
  \begin{axis}[
      nccl axis, 
      xmode=log,
      log basis x=2,
      width=\linewidth,
      height=\linewidth, 
      ymin=0, 
      ymax=90, 
      xmin=0.8, 
      xmax=1280,
      xtick={1, 16, 128, 1024},
      xticklabels={1, 16, 128, 1024},
      ytick={0, 20, 40, 60, 80},
      xlabel={Collective Size (MB)},
      ylabel={P99 CCT (\% of ideal)},
      tick label style={font=\scriptsize},
      label style={font=\footnotesize},
      legend style={
        font=\scriptsize,
        at={(0.02,0.98)}, 
        anchor=north west,
        fill=white,
        fill opacity=0.9,
        draw=gray!50,
      },
  ]

  \addplot[
    color=dark_gray,
    line width=1.75pt,
    mark=*,
    mark size=1.5pt,
    mark options={solid, fill=white, draw=dark_gray, line width=1.2pt},
  ] coordinates {
    (1, 1.194)
    (2, 2.372)
    (4, 4.436)
    (8, 8.288)
    (16, 15.481)
    (32, 26.153)
    (64, 39.898)
    (128, 55.504)
    (256, 69.176)
    (512, 79.053)
    (1024, 85.083)
  };
  \addlegendentry{Flaps}

  \addplot[
    color=nvidia_green,
    line width=1.75pt,
    mark=x,
    mark size=2pt,
    mark options={solid, draw=nvidia_green, line width=1.2pt},
  ] coordinates {
    (1, 1.194)
    (2, 2.372)
    (4, 4.436)
    (8, 8.288)
    (16, 15.481)
    (32, 26.153)
    (64, 39.898)
    (128, 55.504)
    (256, 69.176)
    (512, 79.053)
    (1024, 85.083)
  };
  \addlegendentry{Pristine}

  \end{axis}
\end{tikzpicture}
        \caption{Normalized P99 CCT, 64K nodes, with and without fabric flaps}
        \label{fig:scurve}
    \end{subfigure}
    \hfill 
    \begin{subfigure}[t]{0.48\linewidth} 
        \begin{tikzpicture}[baseline=(current axis.north)]
  \begin{axis}[
      nccl axis, 
      xmode=log,
      width=\linewidth,
      height=\linewidth, 
      ymin=1.0, 
      ymax=3.0, 
      xmin=0.0008, 
      xmax=0.4,
      xtick={0.001, 0.01, 0.1, 0.3},
      xticklabels={0.001, 0.01, 0.1, 0.3},
      ytick={1.0, 1.5, 2.0, 2.5, 3.0},
      xlabel={Convergence Time (s)},
      ylabel={P99 CCT slowdown},
      tick label style={font=\scriptsize},
      label style={font=\footnotesize},
      legend style={
        font=\scriptsize,
        at={(0.02,0.98)}, 
        anchor=north west,
        fill=white,
        fill opacity=0.9,
        draw=gray!50,
      },
  ]

  \addplot[
    color=nvidia_green,
    line width=1.75pt,
    mark=*,
    mark size=1.5pt,
    mark options={solid, fill=nvidia_green},
  ] coordinates {
    (0.001,1.179)
    (0.01,1.207)
    (0.025,1.256)
    (0.05,1.342)
    (0.1,1.533)
    (0.2,1.998)
    (0.3,2.602)
  };

  \end{axis}
\end{tikzpicture}
        \caption{P99 CCT slowdown over pristine, 256K nodes, as function of access link failure convergence}
        \label{fig:exp2}
    \end{subfigure}
    \caption{Large-scale fault tolerance simulation.}
    \label{fig:nsx_failures}
    \vspace{-2pt}
\end{figure}

\Cref{fig:scurve} shows the results normalized by the ideal CCT rate, with and without flaps. The curves overlap, indicating no visible performance impact.

\paragraph{End point link flap} 
We simulate a multi-plane cluster with 256K GPUs and 4 planes. We aim to evaluate the cluster performance as a function of the fabric convergence time, i.e., the time it takes the NIC to converge to the degraded bandwidth over three planes. Until the convergence, we assume that the traffic over the faulty access link gets dropped, which mirrors the real hardware behavior.   We measure P99 CCT among 1024 collectives each 256 ranks.

For this simulation, we consider each NIC to be in one of the three state: pristine (no failed planes), failed (one failed plane before convergence), degraded (one failed plane after convergence to 75\% of the line rate of four planes). We evaluate the latency of a single collective assuming one failed NIC, and record the performance of the collective in the pristine, failed and degraded state.

To evaluate the full workload, we generate failure events in the 256K cluster using the same MTBF as before. Every fault is followed by a convergence event after the convergence time we evaluate, and then a plane restoration event after 10 seconds, matching the flap duration. We run multiple iterations of each set of 1024 collectives, and calculate the collective's performance separately according to the state of the NICs it uses, using the network simulation results. We assume one failure per ring at any point. We estimate P99 CCT over all collectives in the same iteration, and average it over all iterations in the trace.

The MTBF and flap duration are the same as in the fabric. We note that such a  flap rate is very high and was observed only in a small fraction of the cluster time.   

As shown in Figure~\ref{fig:exp2} the P99 CCT increases by 20\% due to plane bandwidth loss, as long as the convergence is within 10 milliseconds. This is inevitable, since there is at least one fault affecting at least one collective. Higher convergence causes CCT increase by 53\% at 100 msec and 260\% at 300 msec, highlighting the crucial importance of fast convergence in large-scale training.

\subsection{Multiplane load balancing}
\label{ssec:multiplane_lb}
\label{ssec:multiplane}
\label{ssec:source-based}

We evaluate the core design decisions behind \projname's Plane Load Balancer (PLB), focusing on how per-plane congestion control (CC) manages network asymmetry, both in load distribution and reaction time, and demonstrating the importance of decoupled control loops (cf. \S\ref{sec:designprinciples}). 


\begin{figure*}[htbp]
  \centering
  \begin{minipage}[b]{0.42\linewidth}
    \centering
    \pgfplotslegendfromname{bwlegend}\\[0.2em]
    \begin{subfigure}[b]{0.49\linewidth}
    \begin{tikzpicture}
      \begin{axis}[
        nccl axis,
        xmode=log,
        width=\linewidth, height=3.0cm,
        xmin=900000, xmax=5.5e9,
        xtick={1048576,16777216,268435456,1073741824,4294967296},
        xticklabels={1\,MB,16\,MB,256\,MB,1\,GB,4\,GB},
        xticklabel style={rotate=40, anchor=east, font=\footnotesize},
        ymin=0, ymax=110,
        ytick={0,25,50,75,100},
        ylabel={Bus bandwidth (GB/s)},
        ylabel style={font=\footnotesize},
        legend to name=bwlegend,
        legend columns=2,
        legend style={
          font=\scriptsize, cells={anchor=west},
          fill=white, fill opacity=0.95, draw=gray!50,
          column sep=0.6em, row sep=0.1em,
        },
      ]
        \addplot[nvidia_green, solid, very thick, mark=o, mark size=1.5pt,
                 mark options={solid, fill=nvidia_green}] coordinates {
          (1048576,5.874)(2097152,11.257)(4194304,20.011)(8388608,32.373)
          (16777216,48.54)(33554432,64.334)(67108864,76.139)(134217728,84.741)
          (268435456,89.373)(536870912,92.028)(1073741824,93.388)
          (2147483648,94.106)(4294967296,94.415)
        };
        \addlegendentry{SPX (baseline)}

        \addplot[nvidia_green, densely dashed, very thick, mark=square, mark size=1.5pt,
                 mark options={solid, fill=nvidia_green}] coordinates {
          (1048576,5.751)(2097152,9.721)(4194304,17.044)(8388608,28.318)
          (16777216,43.1)(33554432,60.896)(67108864,74.518)(134217728,82.823)
          (268435456,87.743)(536870912,90.487)(1073741824,92.358)
          (2147483648,93.293)(4294967296,93.712)
        };
        \addlegendentry{SPX (asymmetry)}

        \addplot[dark_gray, solid, very thick, mark=triangle, mark size=1.5pt,
                 mark options={solid, fill=dark_gray}] coordinates {
          (1048576,6.091)(2097152,11.49)(4194304,20.464)(8388608,32.794)
          (16777216,48.748)(33554432,63.91)(67108864,76.628)(134217728,84.682)
          (268435456,89.451)(536870912,92.148)(1073741824,93.522)
          (2147483648,94.248)(4294967296,94.509)
        };
        \addlegendentry{Global CC (baseline)}

        \addplot[dark_gray, densely dashed, very thick, mark=diamond, mark size=1.5pt,
                 mark options={solid, fill=dark_gray}] coordinates {
          (1048576,4.954)(2097152,6.514)(4194304,10.99)(8388608,14.264)
          (16777216,16.469)(33554432,20.865)(67108864,26.856)(134217728,32.952)
          (268435456,38.023)(536870912,41.199)(1073741824,44.847)
          (2147483648,46.131)(4294967296,47.298)
        };
        \addlegendentry{Global CC (asymmetry)}
      \end{axis}
    \end{tikzpicture}
    \caption{One-to-Many}
    \label{fig:bw_otm}
  \end{subfigure}
  \hfill
  \begin{subfigure}[b]{0.49\linewidth}
    \begin{tikzpicture}
      \begin{axis}[
        nccl axis,
        xmode=log,
        width=\linewidth, height=3.0cm,
        xmin=900000, xmax=5.5e9,
        xtick={1048512,16777152,268435392,1073741760,4294967232},
        xticklabels={1\,MB,16\,MB,256\,MB,1\,GB,4\,GB},
        xticklabel style={rotate=40, anchor=east, font=\footnotesize},
        ymin=0, ymax=110,
        ytick={0,25,50,75,100},
        ylabel={Bus bandwidth (GB/s)},
        ylabel style={font=\footnotesize}
      ]
        \addplot[nvidia_green, solid, very thick, mark=o, mark size=1.5pt,
                 mark options={solid, fill=nvidia_green}] coordinates {
          (1048512,2.835)(2097120,5.913)(4194240,9.818)(8388576,14.8)
          (16777152,26.622)(33554400,33.598)(67108800,38.497)(134217696,64.468)
          (268435392,79.9)(536870880,86.152)(1073741760,89.377)
          (2147483616,91.104)(4294967232,91.944)
        };
        \addplot[nvidia_green, densely dashed, very thick, mark=square, mark size=1.5pt,
                 mark options={solid, fill=nvidia_green}] coordinates {
          (1048512,2.679)(2097120,4.761)(4194240,7.788)(8388576,11.041)
          (16777152,21.566)(33554400,29.798)(67108800,35.77)(134217696,60.325)
          (268435392,71.969)(536870880,78.88)(1073741760,83.979)
          (2147483616,87.325)(4294967232,88.987)
        };
        \addplot[dark_gray, solid, very thick, mark=triangle, mark size=1.5pt,
                 mark options={solid, fill=dark_gray}] coordinates {
          (1048512,2.885)(2097120,5.805)(4194240,9.696)(8388576,14.69)
          (16777152,27.233)(33554400,33.304)(67108800,37.825)(134217696,64.516)
          (268435392,78.84)(536870880,86.026)(1073741760,89.62)
          (2147483616,91.032)(4294967232,92.181)
        };
        \addplot[dark_gray, densely dashed, very thick, mark=diamond, mark size=1.5pt,
                 mark options={solid, fill=dark_gray}] coordinates {
          (1048512,2.656)(2097120,4.764)(4194240,7.237)(8388576,10.827)
          (16777152,18.237)(33554400,23.733)(67108800,28.404)(134217696,35.34)
          (268435392,41.672)(536870880,47.974)(1073741760,52.183)
          (2147483616,54.983)(4294967232,55.855)
        };
      \end{axis}
    \end{tikzpicture}
    \caption{All-to-All}
    \label{fig:bw_a2a}
  \end{subfigure}
  \end{minipage}
  \hfill
  \begin{subfigure}[b]{0.26\linewidth}
    \begin{tikzpicture}
      \begin{axis}[
        nccl axis,
        xmode=log,
        width=\linewidth, height=3.5cm,
        xmin=900000, xmax=5.5e9,
        xtick={1048576,16777216,268435456,1073741824,4294967296},
        xticklabels={1\,MB,16\,MB,256\,MB,1\,GB,4\,GB},
        xticklabel style={rotate=40, anchor=east, font=\footnotesize},
        ymin=0.25, ymax=1.05,
        ytick={0.4,0.6,0.8,1.0},
        ylabel={BW ratio (asym.\ / baseline)},
        ylabel style={font=\footnotesize},
        legend style={font=\scriptsize, at={(0.97,0.03)}, anchor=south east,
                      cells={anchor=west}, fill=white, draw=gray!50},
      ]
        \addplot[nvidia_green, solid, very thick, mark=o, mark size=1.5pt,
                 mark options={solid, fill=nvidia_green}] coordinates {
          (1048576,0.979060)(2097152,0.863552)(4194304,0.851732)(8388608,0.874741)
          (16777216,0.887927)(33554432,0.946560)(67108864,0.978710)(134217728,0.977366)
          (268435456,0.981762)(536870912,0.983255)(1073741824,0.988971)
          (2147483648,0.991361)(4294967296,0.992554)
        };
        \addlegendentry{One-to-Many}

        \addplot[nvidia_green!55!black, solid, very thick, mark=square, mark size=1.5pt,
                 mark options={solid, fill=nvidia_green!55!black}] coordinates {
          (1048512,0.944974)(2097120,0.805175)(4194240,0.793237)(8388576,0.746014)
          (16777152,0.810082)(33554400,0.886898)(67108800,0.929163)(134217696,0.935736)
          (268435392,0.900738)(536870880,0.915591)(1073741760,0.939604)
          (2147483616,0.958520)(4294967232,0.967839)
        };
        \addlegendentry{All-to-All}

        \draw[gray, densely dashed, line width=0.9pt]
          (axis cs:33554432, 0.25) -- (axis cs:33554432, 1.05);
        \node[font=\scriptsize, anchor=north west, xshift=2pt]
          at (axis cs:33554432, 0.82) {${\sim}335\,\mu$s};
      \end{axis}
    \end{tikzpicture}
    \caption{Normalized SPX BW}
    \label{fig:ratio}
  \end{subfigure}
  \hfill
  \begin{subfigure}[b]{0.26\linewidth}
    \input{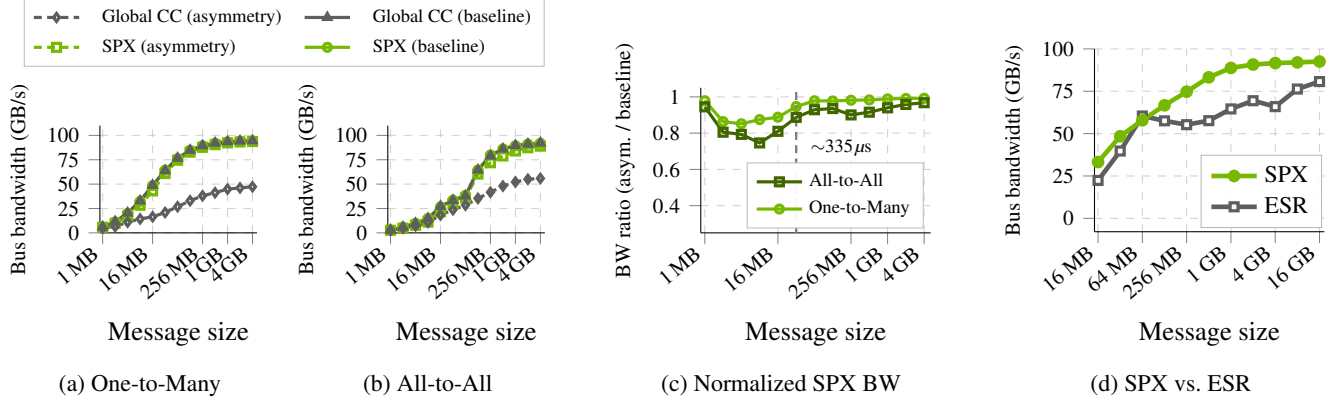}
    \caption{\projname vs.\ ESR}
    \label{fig:sbr_spx_ew}
  \end{subfigure}

  \caption{Collective bandwidth under noise-induced asymmetry (SPX green, Global~CC gray;
    solid: symmetric baseline, dashed: asymmetry); normalized SPX bandwidth ratio
    (dashed line: ${\sim}335\,\mu$s convergence);
     \ATOA  \projname vs.\ Entropy Source Routing.}
  \label{fig:bw}
\end{figure*}
\input{figures/mp_asymmetry_setup}

\paragraph{No degradation under dynamic asymmetry} 
Dynamic asymmetry occurs when some plane(s) becomes degraded and gets saturated, whereas other planes still have spare capacity.  Such scenarios are challenging for load balancing. 

We configure a 16-NIC-per-leaf subset of the Blackwell\_Ultra\_MP cluster (\cref{fig:mp-asymmetry-setup}), trimming the uplinks on Leaf~2 (plane 2) and Leaf~3 (plane 3) from a non-blocking 16$\times$200\,Gbps to 4$\times$200\,Gbps. We run the main workload on 8~NICs per leaf and noise on the remaining 8~NICs.

We execute the main workload in two modes: \emph{one-to-many} (NICs under Leaf~1 send repeated bursts to hosts under Leaf~2 and Leaf~3) and \emph{\ATOA} executed by all NICs. Both workloads stress the separation of control loops: the CC must throttle per-destination rates under incast pressure, while the PLB must simultaneously divert traffic away from the degraded planes. These two reactions must proceed independently to avoid interfering with each other's signals. 

We establish a baseline by executing the tests without asymmetry, and compare \projname PLB against \textbf{Global~CC}, a variant using a single shared CC context across all planes.

\Cref{fig:bw} shows the results.
Under uniform fabric \projname and Global~CC perform identically,
confirming that per-plane state introduces no overhead. Under asymmetry, however, they diverge: without per-plane CC, Global~CC's bandwidth collapses from 94.5\,GB/s to 47.3\,GB/s (a $>50\%$ drop) for the one-to-many workload, and from 92.2\,GB/s to 55.9\,GB/s (a $>40\%$ drop) for \ATOA. In contrast, \projname isolates the congestion, sustaining 93.7\,GB/s (one-to-many) and 89.0\,GB/s (\ATOA) for 4\,GB messages. We see near-baseline performance despite the 75\% uplink reduction on two leaf switches.

\paragraph{Fast convergence within critical message size ranges}
\Cref{fig:ratio} plots \projname's normalized bandwidth under dynamic asymmetry. For messages below ${\sim}$4\,MB, the PLB has not yet accumulated sufficient per-plane congestion signals to detect the imbalance, yielding normalized performance ratios as low as 0.85 for one-to-many and 0.75 for \ATOA. Since the background noise is continuous, the PLB state resets between bursts and must re-learn at each iteration. However, for message sizes above ${\sim}$335\,$\mu$s threshold (32\,MB), PLB fully converges, allowing \projname to sustain $>97\%$ and $>90\%$ of its baseline bandwidth for the one-to-many and \ATOA workloads, respectively.

\paragraph{Conflicting control loops: entropy-based source routing} We assess the importance of decoupled control loops by comparing \projname's PLB to \textbf{entropy-based source routing (ESR)}. In ESR, entropy values jointly encode both the target plane and the intra-plane path, fundamentally entangling congestion control (CC) and load balancing. As a result, the aggregate CC state cannot provide independent steering signals for plane selection and intra-plane routing. We measure the impact of this entanglement by running four concurrent \ATOA collectives across 32 nodes with NVLink disabled (scale-out network traffic only). As shown in \Cref{fig:sbr_spx_ew}, \projname scales smoothly to ${\sim}$92\,GB/s. In contrast, ESR exhibits oscillating throughput (55--80\,GB/s above 256\,MB) due to its conflicting control loops. These results confirm that per-plane CC design is beneficial for multiplane load balancing.


\section{Related work}
\label{sec:related_work}

\paragraph{Production large-scale RoCE and AI fabrics}
This paper introduces  a novel hardware-accelerated multi-plane architecture deployed in production large-scale clusters, complementing prior works on RDMA-over-Ethernet and AI-training fabrics at hyperscale, including Meta~\cite{meta_roce_lossless}, Microsoft~\cite{microsoft_roce_lossless} and Alibaba~\cite{alibaba_hpn,stellar-sigcomm25}, and Tencent's Astral with its same-rail tier-2 interconnect for half-a-million-GPU training~\cite{astral-sigcomm25}.

\paragraph{Multi-path RDMA transport and packet spraying}
A parallel line of work pursues multi-path load balancing at the transport
layer. The Ultra Ethernet Consortium's UET
specification~\cite{uec_uet_spec} standardizes entropy-based per-packet
spraying,  Google's PLB~\cite{plb-sigcomm22} uses host-side congestion signals for exploiting multiple paths. 
 These schemes react at multi-RTT, host-controlled
timescales and typically maintain a single congestion-control loop with no
per-path visibility. \projname takes a different approach of hardware-accelerated in-network load balancing and control loop separation. 

\paragraph{In-network load balancing and adaptive routing}
Fine-grained in-network load balancing has been studied extensively~\cite{Adaptive_routing,drill-sigcomm17, conga-sigcomm14,letflow-nsdi17,conweave-sigcomm23,sglb-sigcomm25},
%
%
Weighted multi-path
forwarding under capacity asymmetry has been addressed by
WCMP~\cite{wcmp-eurosys14} and, more recently, by Juniper's weighted packet
spray for AI/ML fabrics~\cite{juniper_weighted_packet_spray,juniper_ai_dc_wp},
which bias per-packet forwarding by static link weights but do not specify
composing those weights with queue-aware adaptive port selection.
\projname introduces many novel components, taking adaptive routing to extreme with hardware-accelerated multi-plane design.


\paragraph{Congestion control}
\projname builds on a well-established RDMA CC design space,
including ECN-based rate control (DCQCN~\cite{dcqcn}), delay-based signaling
(Swift~\cite{swift_google}), in-network telemetry (HPCC~\cite{hpcc-sigcomm19}),
and revisited loss-recovery assumptions (IRN~\cite{mittal2018revisiting}).
We show how to combine these with adaptive routing to optimize for AI training workloads,  keeping the control loop tuned to sustained collective congestion
rather than transient micro-bursts that can be balanced in the fabric.

\section{Conclusions}
\label{sec:discussion}


We described Spectrum-X, a field-tested Ethernet fabric running in multiple large-scale AI training clusters. Following our rich deployment experience, four takeaways stand out.
First, topology matters as much as transport. Given the growing demand for scale, building shallow multiplane topologies with high-radix switches is key to achieving better clusters across every critical dimension: performance, fault tolerance, power, cost, and cabling complexity. 
%
%
Second, hardware acceleration with clear signal and control loop separation is what makes the fabric work at 800 Gbps and beyond. Host-based or software-driven alternatives failed in our evaluations to provide the convergence time that synchronous collectives demand.
%
%
Third, operating a fabric at this scale depends on the visibility and ease-of-tuning.
%
Last, validating cluster and networking technology at this scale requires a well-defined set of benchmarks with clear KPIs that can be run consistently across production clusters, proxy-scale testbeds, and simulators; defining and standardizing this benchmark set is, in our view, a prerequisite for healthy progress in the field.


\section*{Acknowledgments}
We thank the anonymous reviewers for their insightful 
comments. We also thank our colleagues at NVIDIA, including those in the Networking Software, Networking Research, Networking Software \& Systems Architecture, and Networking Product Marketing organizations, for their contributions to the development of NVIDIA Spectrum-X.

\bibliographystyle{plain}
\balance
\bibliography{verified_networking_papers}

\begin{thebibliography}{10}

\bibitem{conga-sigcomm14}
Mohammad Alizadeh, Tom Edsall, Sarang Dharmapurikar, Ramanan Vaidyanathan, Kevin Chu, Andy Fingerhut, Vinh~The Lam, Francis Matus, Rong Pan, Navindra Yadav, and George Varghese.
\newblock {CONGA: Distributed Congestion-Aware Load Balancing for Datacenters}.
\newblock In {\em ACM SIGCOMM}, 2014.

\bibitem{blog_spx_storage}
Taylor Allison.
\newblock {Accelerating {AI} Storage by up to 48\% with {NVIDIA} {Spectrum-X} Networking Platform and Partners}.
\newblock \url{https://developer.nvidia.com/blog/accelerating-ai-storage-by-up-to-48-with-nvidia-spectrum-x-networking-platform-and-partners/}, 2025.

\bibitem{microsoft_roce_lossless}
Wei Bai, Shanim~Sainul Abdeen, Ankit Agrawal, Krishan~Kumar Attre, Paramvir Bahl, Ameya Bhagat, Gowri Bhaskara, Tanya Brokhman, Lei Cao, Ahmad Cheema, Rebecca Chow, Jeff Cohen, Mahmoud Elhaddad, Vivek Ette, Igal Figlin, Daniel Firestone, Mathew George, Ilya German, Lakhmeet Ghai, Eric Green, Albert Greenberg, Manish Gupta, Randy Haagens, Matthew Hendel, Ridwan Howlader, Neetha John, Julia Johnstone, Tom Jolly, Greg Kramer, David Kruse, Ankit Kumar, Erica Lan, Ivan Lee, Avi Levy, Marina Lipshteyn, Xin Liu, Chen Liu, Guohan Lu, Yuemin Lu, Xiakun Lu, Vadim Makhervaks, Ulad Malashanka, David~A. Maltz, Ilias Marinos, Rohan Mehta, Sharda Murthi, Anup Namdhari, Aaron Ogus, Jitendra Padhye, Madhav Pandya, Douglas Phillips, Adrian Power, Suraj Puri, Shachar Raindel, Jordan Rhee, Anthony Russo, Maneesh Sah, Ali Sheriff, Chris Sparacino, Ashutosh Srivastava, Weixiang Sun, Nick Swanson, Fuhou Tian, Lukasz Tomczyk, Vamsi Vadlamuri, Alec Wolman, Ying Xie, Joyce Yom, Lihua Yuan, Yanzhao Zhang, and Brian Zill.
\newblock {Empowering Azure Storage with {RDMA}}.
\newblock In {\em USENIX NSDI}, 2023.

\bibitem{deepseek-v3}
DeepSeek-AI.
\newblock {DeepSeek-V3 Technical Report}, 2024.

\bibitem{hpcwire_colossus}
Doug Eadline.
\newblock {xAI Colossus: The Elon Project}.
\newblock \url{https://www.hpcwire.com/2024/09/05/xai-colossus-the-elon-project/}, 2024.

\bibitem{meta_roce_lossless}
Adithya Gangidi, Rui Miao, Shengbao Zheng, Sai~Jayesh Bondu, Guilherme Goes, Hany Morsy, Rohit Puri, Mohammad Riftadi, Ashmitha~Jeevaraj Shetty, Jingyi Yang, Shuqiang Zhang, Mikel~Jimenez Fernandez, Shashidhar Gandham, and Hongyi Zeng.
\newblock {RDMA over Ethernet for Distributed Training at Meta Scale}.
\newblock In {\em ACM SIGCOMM}, 2024.

\bibitem{drill-sigcomm17}
Soudeh Ghorbani, Zibin Yang, P.~Brighten Godfrey, Yashar Ganjali, and Amin Firoozshahian.
\newblock {DRILL: Micro Load Balancing for Low-latency Data Center Networks}.
\newblock In {\em ACM SIGCOMM}, 2017.

\bibitem{juniper_ai_dc_wp}
{Juniper Networks}.
\newblock {Networking the AI Data Center: Advanced Load Balancing ({DLB}, {GLB}, Weighted {ECMP}) for {AI/ML} Fabrics}.
\newblock \url{https://www.juniper.net/content/dam/www/assets/white-papers/us/en/networking-the-ai-data-center.pdf}, 2024.

\bibitem{juniper_weighted_packet_spray}
{Juniper Networks}.
\newblock {Weighted Packet Spray for Dynamic Load Balancing ({Junos} OS Evolved {AI/ML} Guide)}.
\newblock \url{https://www.juniper.net/documentation/us/en/software/junos/ai-ml-evo/topics/topic-map/weighted-packet-spray.html}, 2024.

\bibitem{nsx}
Sajy Khashab, Hariharan Sezhiyan, Rani Abboud, Alex Normatov, Stefan Kaestle, Eliav Bar-Ilan, Mohammad Nassar, Omer Shabtai, Wei Bai, Matty Kadosh, Jiarong Xing, Mark Silberstein, T.~S.~Eugene Ng, and Ang Chen.
\newblock {NSX: Large-Scale Network Simulation on an AI Server}.
\newblock In {\em ACM NAIC}, 2025.

\bibitem{swift_google}
Gautam Kumar, Nandita Dukkipati, Keon Jang, Hassan M.~G. Wassel, Xian Wu, Behnam Montazeri, Yaogong Wang, Kevin Springborn, Christopher Alfeld, Michael Ryan, David Wetherall, and Amin Vahdat.
\newblock {Swift: Delay is Simple and Effective for Congestion Control in the Datacenter}.
\newblock In {\em ACM SIGCOMM}, 2020.

\bibitem{hpcc-sigcomm19}
Yuliang Li, Rui Miao, Hongqiang~Harry Liu, Yan Zhuang, Fei Feng, Lingbo Tang, Zheng Cao, Ming Zhang, Frank Kelly, Mohammad Alizadeh, and Minlan Yu.
\newblock {HPCC: High Precision Congestion Control}.
\newblock In {\em ACM SIGCOMM}, 2019.

\bibitem{lin2025understandingstragglerslargemodel}
Jinkun Lin, Ziheng Jiang, Zuquan Song, Sida Zhao, Menghan Yu, Zhanghan Wang, Chenyuan Wang, Zuocheng Shi, Xiang Shi, Wei Jia, Zherui Liu, Shuguang Wang, Haibin Lin, Xin Liu, Aurojit Panda, and Jinyang Li.
\newblock {Understanding stragglers in large model training using what-if analysis}.
\newblock In {\em USENIX OSDI}, 2025.

\bibitem{ietf-bess-ebgp-dmz-10}
Stephane Litkowski, SATYA~R MOHANTY, Arie Vayner, Akshay Gattani, Ajay Kini, Jeff Tantsura, and Reshma Das.
\newblock {BGP link bandwidth extended community use cases}.
\newblock Internet-Draft draft-ietf-bess-ebgp-dmz-10, 2026.

\bibitem{stellar-sigcomm25}
Jie Lu, Jiaqi Gao, Fei Feng, Zhiqiang He, Menglei Zheng, Kun Liu, Jun He, Binbin Liao, Suwei Xu, Ke~Sun, Yongjia Mo, Qinghua Peng, Jilie Luo, Qingxu Li, Gang Lu, Zishu Wang, Jianbo Dong, Kunling He, Sheng Cheng, Jiamin Cao, Hairong Jiao, Pengcheng Zhang, Shu Ma, Lingjun Zhu, Chao Shi, Yangming Zhang, Yiquan Chen, Wei Wang, Shuhong Zhu, Xingru Li, Qiang Wang, Jiang Liu, Chao Wang, Wei Lin, Ennan Zhai, Jiesheng Wu, Qiang Liu, Binzhang Fu, and Dennis Cai.
\newblock {Alibaba Stellar: A New Generation RDMA Network for Cloud AI}.
\newblock In {\em ACM SIGCOMM}, 2025.

\bibitem{mcclure2025load}
Sarah McClure, Evyatar Cohen, Alex Shpiner, Mark Silberstein, Sylvia Ratnasamy, Scott Shenker, and Isaac Keslassy.
\newblock {Load Balancing for AI Training Workloads}, 2026.

\bibitem{astral-sigcomm25}
Qingkai Meng, Hao Zheng, Zhenhui Zhang, ChonLam Lao, Chengyuan Huang, Baojia Li, Ziyuan Zhu, Hao Lu, Weizhen Dang, Zitong Lin, Weifeng Zhang, Lingfeng Liu, Yuanyuan Gong, Chunzhi He, Xiaoyuan Hu, Yinben Xia, Xiang Li, Zekun He, Yachen Wang, Xianneng Zou, Kun Yang, Gianni Antichi, Guihai Chen, and Chen Tian.
\newblock {Astral: A Datacenter Infrastructure for Large Language Model Training at Scale}.
\newblock In {\em ACM SIGCOMM}, 2025.

\bibitem{mittal2018revisiting}
Radhika Mittal, Alexander Shpiner, Aurojit Panda, Eitan Zahavi, Arvind Krishnamurthy, Sylvia Ratnasamy, and Scott Shenker.
\newblock {Revisiting network support for RDMA}.
\newblock In {\em ACM SIGCOMM}, 2018.

\bibitem{swarm-nsdi25}
Pooria Namyar, Arvin Ghavidel, Daniel Crankshaw, Daniel~S. Berger, Kevin Hsieh, Srikanth Kandula, Ramesh Govindan, and Behnaz Arzani.
\newblock {Enhancing Network Failure Mitigation with Performance-Aware Ranking}.
\newblock In {\em USENIX NSDI}, 2025.

\bibitem{spx_h100_announcement}
{NVIDIA}.
\newblock {NVIDIA Launches Accelerated Ethernet Platform for Hyperscale Generative AI (Spectrum-X)}.
\newblock \url{https://nvidianews.nvidia.com/news/nvidia-launches-accelerated-ethernet-platform-for-hyperscale-generative-ai}, 2023.

\bibitem{nccl_docs}
{NVIDIA}.
\newblock {NCCL Developer Guide and Environment Variables}.
\newblock \url{https://docs.nvidia.com/deeplearning/nccl/user-guide/docs/env.html}, 2024.

\bibitem{nccl_tests_perf}
{NVIDIA}.
\newblock {NCCL Tests: Performance --- Bus Bandwidth Metric}.
\newblock \url{https://github.com/NVIDIA/nccl-tests/blob/master/doc/PERFORMANCE.md}, 2024.

\bibitem{blog_xgs}
{NVIDIA}.
\newblock {{NVIDIA} Introduces {Spectrum-XGS} Ethernet to Connect Distributed Data Centers Into Giga-Scale {AI} Super-Factories}.
\newblock \url{https://nvidianews.nvidia.com/news/nvidia-introduces-spectrum-xgs-ethernet-to-connect-distributed-data-centers-into-giga-scale-ai-super-factories}, 2025.

\bibitem{spx_nvidia_website}
{NVIDIA}.
\newblock {{NVIDIA} {Spectrum-X} Networking Platform}.
\newblock \url{https://www.nvidia.com/en-us/networking/spectrumx/}, 2026.

\bibitem{sglb-sigcomm25}
Chenchen Qi, Wenfei Wu, Yongcan Wang, Keqiang He, Yu-Hsiang Kao, Zongying He, Chen-Yu Yen, Zhuo Jiang, Feng Luo, Surendra Anubolu, Yanjin Gao, Bingfeng Lin, Wenda Ni, Yiming Yang, Donglin Wei, Boyang Zhou, Jian Wang, and Shan Ding.
\newblock {SGLB: Scalable and Robust Global Load Balancing in Commodity AI Clusters}.
\newblock In {\em ACM SIGCOMM}, 2025.

\bibitem{alibaba_hpn}
Kun Qian, Yongqing Xi, Jiamin Cao, Jiaqi Gao, Yichi Xu, Yu~Guan, Binzhang Fu, Xuemei Shi, Fangbo Zhu, Rui Miao, Chao Wang, Peng Wang, Pengcheng Zhang, Xianlong Zeng, Eddie Ruan, Zhiping Yao, Ennan Zhai, and Dennis Cai.
\newblock {Alibaba HPN: A Data Center Network for Large Language Model Training}.
\newblock In {\em ACM SIGCOMM}, 2024.

\bibitem{plb-sigcomm22}
Mubashir~Adnan Qureshi, Yuchung Cheng, Qianwen Yin, Qiaobin Fu, Gautam Kumar, Masoud Moshref, Junhua Yan, Van Jacobson, David Wetherall, and Abdul Kabbani.
\newblock {PLB: Congestion Signals are Simple and Effective for Network Load Balancing}.
\newblock In {\em ACM SIGCOMM}, 2022.

\bibitem{pronto-nsdi17}
Arjun Roy, Hongyi Zeng, Jasmeet Bagga, and Alex~C. Snoeren.
\newblock {Passive Realtime Datacenter Fault Detection and Localization}.
\newblock In {\em USENIX NSDI}, 2017.

\bibitem{blog_ns}
Shashank Sabhlok.
\newblock {North--South Networks: The Key to Faster Enterprise {AI} Workloads}.
\newblock \url{https://developer.nvidia.com/blog/north-south-networks-the-key-to-faster-enterprise-ai-workloads/}, 2025.

\bibitem{conweave-sigcomm23}
Cha~Hwan Song, Xin~Zhe Khooi, Raj Joshi, Inho Choi, Jialin Li, and Mun~Choon Chan.
\newblock {Network Load Balancing with In-network Reordering Support for RDMA}.
\newblock In {\em ACM SIGCOMM}, 2023.

\bibitem{sonic_pfc_watchdog}
{SONiC Project}.
\newblock {PFC} watchdog design.
\newblock \url{https://github.com/sonic-net/SONiC/wiki/PFC-Watchdog-Design}, 2018.

\bibitem{netbouncer-nsdi19}
Cheng Tan, Ze~Jin, Chuanxiong Guo, Tianrong Zhang, Haitao Wu, Karl Deng, Dongming Bi, and Dong Xiang.
\newblock {NetBouncer: Active Device and Link Failure Localization in Data Center Networks}.
\newblock In {\em USENIX NSDI}, 2019.

\bibitem{uec_uet_spec}
{Ultra Ethernet Consortium}.
\newblock {Ultra Ethernet Specification v1.0}.
\newblock \url{https://ultraethernet.org/wp-content/uploads/sites/20/2025/06/UE-Specification-6.11.25.pdf}, 2025.

\bibitem{letflow-nsdi17}
Erico Vanini, Rong Pan, Mohammad Alizadeh, Parvin Taheri, and Tom Edsall.
\newblock {Let It Flow: Resilient Asymmetric Load Balancing with Flowlet Switching}.
\newblock In {\em USENIX NSDI}, 2017.

\bibitem{optireduce-nsdi25}
Ertza Warraich, Omer Shabtai, Khalid Manaa, Shay Vargaftik, Yonatan Piasetzky, Matty Kadosh, Lalith Suresh, and Muhammad Shahbaz.
\newblock {OptiReduce: Resilient and Tail-Optimal AllReduce for Distributed Deep Learning in the Cloud}.
\newblock In {\em USENIX NSDI}, 2025.

\bibitem{xai_colossus}
{xAI}.
\newblock {Colossus: The World's Largest {AI} Supercomputer}.
\newblock \url{https://x.ai/colossus}, 2024.

\bibitem{holmes-nsdi25}
Zhiyi Yao, Pengbo Hu, Congcong Miao, Xuya Jia, Zuning Liang, Yuedong Xu, Chunzhi He, Hao Lu, Mingzhuo Chen, Xiang Li, Zekun He, Yachen Wang, Xianneng Zou, and Junchen Jiang.
\newblock {Holmes: Localizing Irregularities in {LLM} Training with Mega-scale {GPU} Clusters}.
\newblock In {\em USENIX NSDI}, 2025.

\bibitem{Adaptive_routing}
Eitan Zahavi, Isaac Keslassy, and Avinoam Kolodny.
\newblock {Distributed adaptive routing for big-data applications running on data center networks}.
\newblock In {\em ACM/IEEE ANCS}, 2012.

\bibitem{zhao-nsdi19-rewiring}
Shizhen Zhao, Rui Wang, Junlan Zhou, Joon Ong, Jeffrey~C. Mogul, and Amin Vahdat.
\newblock {Minimal Rewiring: Efficient Live Expansion for {Clos} Data Center Networks}.
\newblock In {\em USENIX NSDI}, 2019.

\bibitem{wcmp-eurosys14}
Junlan Zhou, Malveeka Tewari, Min Zhu, Abdul Kabbani, Leon Poutievski, Arjun Singh, and Amin Vahdat.
\newblock {WCMP: Weighted Cost Multipathing for Improved Fairness in Data Centers}.
\newblock In {\em EuroSys}, 2014.

\bibitem{dcqcn}
Yibo Zhu, Haggai Eran, Daniel Firestone, Chuanxiong Guo, Marina Lipshteyn, Yehonatan Liron, Jitendra Padhye, Shachar Raindel, Mohamad~Haj Yahia, and Ming Zhang.
\newblock {Congestion Control for Large-Scale RDMA Deployments}.
\newblock In {\em ACM SIGCOMM}, 2015.

\end{thebibliography}


\end{document}